\documentclass[11pt,oneside]{book}
\usepackage[T2A,T1]{fontenc}
\usepackage[utf8]{inputenc}
\usepackage{microtype}
\usepackage[russian,english]{babel}

\usepackage[within=none,figurename=Fig.]{caption} 
\makeatletter\renewcommand\@biblabel[1]{#1.}\makeatother 
\usepackage{etoolbox}
\patchcmd{\thebibliography}{\chapter*}{\section*}{}{}

\usepackage{amsmath,amsfonts,amssymb,amsthm}

\newtheoremstyle{nofullstop} 
  {\topsep}{\topsep}%
  {\itshape}{}%
  {\bfseries}{}{5pt plus 1pt minus 1pt}{}%
\theoremstyle{nofullstop}
\newtheorem{thm}{Theorem}[section]

\renewcommand\emptyset{\varnothing} 
\renewcommand\epsilon{\varepsilon} 

\newcommand{\poly}{\operatorname{poly}}
\newcommand{\col}{f}

\usepackage{txfonts}
\usepackage[T1]{fontenc}
\usepackage{helvet}

\usepackage{booktabs}

\usepackage{xcolor}
\definecolor{col1}{RGB}{255,255,255}
\definecolor{col2}{RGB}{200,200,200}
\definecolor{col3}{RGB}{150,150,150}
\definecolor{col4}{RGB}{100,100,100}
\definecolor{col5}{RGB}{50,50,50}

\definecolor{lightgray}{RGB}{180,180,180}

\usepackage{tikz}
\usetikzlibrary{chains, calc, backgrounds,patterns}
\tikzset{v/.style={draw,fill,circle, inner sep=1pt}}

\newcommand{\women}{
\node  (Flora Price) at (17,16.5) {};
\node  (Nora Fayette) at (9,21) {};
\node (E10) at (2,18) {};
\node  (Verne Sanderson) at (16,14.5) {};
\node (E12) at (11,19.5) {};
\node (E14) at (3,20.5) {};
\node  (Frances Anderson) at (2,6) {};
\node  (Dorothy Murchison) at (8,15.5) {};
\node  (Evelyn Jefferson) at (16,9) {};
\node  (Ruth DeSand) at (16,10.5) {};
\node  (Helen Lloyd) at (11.6,16.0) {};
\node  (Olivia Carleton) at (15.5,17.3) {};
\node  (Eleanor Nye) at (1.6,9.6) {};
\node (E11) at (16,20) {};
\node (E9) at (11.1,13.2) {};
\node (E8) at (5.5,13.5) {};
\node (E5) at (5,6) {};
\node (E4) at (13,2.5) {};
\node (E7) at (8.8,9.9) {};
\node (E6) at (5.9,8.1) {};
\node (E1) at (19.1,6.8) {};
\node  (Myra Liddel) at (1.9,15.3) {};
\node (E3) at (7.4,2.1) {};
\node (E2) at (17.7,4.3) {};
\node  (Theresa Anderson) at (12.2,9.8) {};
\node  (Pearl Oglethorpe) at (5.5,11.2) {};
\node  (Katherina Rogers) at (7.5,18.7) {};
\node  (Brenda Rogers) at (13.6,6.4) {};
\node (E13) at (5.5,22.0) {};
\node  (Charlotte McDowd) at (7.4,4.3) {};
\node  (Sylvia Avondale) at (5.5,16.0) {};
\node  (Laura Mandeville) at (10.4,5.0) {};
\draw (Flora Price)--(E11);
\draw (Flora Price)--(E9);
\draw (Nora Fayette)--(E11);
\draw (Nora Fayette)--(E10);
\draw (Nora Fayette)--(E13);
\draw (Nora Fayette)--(E12);
\draw (Nora Fayette)--(E14);
\draw (Nora Fayette)--(E9);
\draw (Nora Fayette)--(E7);
\draw (Nora Fayette)--(E6);
\draw (E10)--(Sylvia Avondale);
\draw (E10)--(Myra Liddel);
\draw (E10)--(Helen Lloyd);
\draw (E10)--(Katherina Rogers);
\draw (Verne Sanderson)--(E7);
\draw (Verne Sanderson)--(E12);
\draw (Verne Sanderson)--(E9);
\draw (Verne Sanderson)--(E8);
\draw (E12)--(Helen Lloyd);
\draw (E12)--(Katherina Rogers);
\draw (E12)--(Sylvia Avondale);
\draw (E12)--(Myra Liddel);
\draw (E14)--(Sylvia Avondale);
\draw (E14)--(Katherina Rogers);
\draw (Frances Anderson)--(E5);
\draw (Frances Anderson)--(E6);
\draw (Frances Anderson)--(E8);
\draw (Frances Anderson)--(E3);
\draw (Dorothy Murchison)--(E9);
\draw (Dorothy Murchison)--(E8);
\draw (Evelyn Jefferson)--(E9);
\draw (Evelyn Jefferson)--(E8);
\draw (Evelyn Jefferson)--(E5);
\draw (Evelyn Jefferson)--(E4);
\draw (Evelyn Jefferson)--(E6);
\draw (Evelyn Jefferson)--(E1);
\draw (Evelyn Jefferson)--(E3);
\draw (Evelyn Jefferson)--(E2);
\draw (Ruth DeSand)--(E5);
\draw (Ruth DeSand)--(E7);
\draw (Ruth DeSand)--(E9);
\draw (Ruth DeSand)--(E8);
\draw (Helen Lloyd)--(E11);
\draw (Helen Lloyd)--(E7);
\draw (Helen Lloyd)--(E8);
\draw (Olivia Carleton)--(E11);
\draw (Olivia Carleton)--(E9);
\draw (Eleanor Nye)--(E5);
\draw (Eleanor Nye)--(E7);
\draw (Eleanor Nye)--(E6);
\draw (Eleanor Nye)--(E8);
\draw (E9)--(Theresa Anderson);
\draw (E9)--(Pearl Oglethorpe);
\draw (E9)--(Katherina Rogers);
\draw (E9)--(Sylvia Avondale);
\draw (E9)--(Myra Liddel);
\draw (E8)--(Theresa Anderson);
\draw (E8)--(Pearl Oglethorpe);
\draw (E8)--(Katherina Rogers);
\draw (E8)--(Sylvia Avondale);
\draw (E8)--(Brenda Rogers);
\draw (E8)--(Laura Mandeville);
\draw (E8)--(Myra Liddel);
\draw (E5)--(Theresa Anderson);thick,
\draw (E5)--(Brenda Rogers);
\draw (E5)--(Laura Mandeville);
\draw (E5)--(Charlotte McDowd);
\draw (E4)--(Charlotte McDowd);
\draw (E4)--(Theresa Anderson);
\draw (E4)--(Brenda Rogers);
\draw (E7)--(Theresa Anderson);
\draw (E7)--(Sylvia Avondale);
\draw (E7)--(Brenda Rogers);
\draw (E7)--(Laura Mandeville);
\draw (E7)--(Charlotte McDowd);
\draw (E6)--(Theresa Anderson);
\draw (E6)--(Pearl Oglethorpe);
\draw (E6)--(Brenda Rogers);
\draw (E6)--(Laura Mandeville);
\draw (E1)--(Laura Mandeville);
\draw (E1)--(Brenda Rogers);
\draw (E3)--(Theresa Anderson);
\draw (E3)--(Brenda Rogers);
\draw (E3)--(Laura Mandeville);
\draw (E3)--(Charlotte McDowd);
\draw (E2)--(Laura Mandeville);
\draw (E2)--(Theresa Anderson);
\draw (Katherina Rogers)--(E13);
\draw (E13)--(Sylvia Avondale);}
\newcommand{\florentineee}{
  \node (Acciaiuoli) [v] at (0.41, 0.77) {};
  \node (Albizzi) [v] at (0.20, 0.36) {};
  \node (Barbadori) [v] at (0.62, 0.50) {};
  \node (Bischeri) [v] at (0.55, 0.12) {};
  \node (Castellani) [v] at (0.75, 0.34) {};
  \node (Ginori) [v] at (0.00, 0.35) {};
  \node (Guadagni) [v] at (0.32, 0.18) {};
  \node (Lamberteschi) [v] at (0.22, 0.00) {};
  \node (Medici) [v] at (0.38, 0.54) {};
  \node (Pazzi) [v] at (0.10, 0.75) {};
  \node (Peruzzi) [v] at (0.73, 0.16) {};
  \node (Pucci) [v] at (0.7, 0.7) {};
  \node (Ridolfi) [v] at (0.50, 0.39) {};
  \node (Salviati) [v] at (0.24, 0.65) {};
  \node (Strozzi) [v] at (0.61, 0.28) {};
  \node (Tornabuoni) [v] at (0.36, 0.36) {};
    \draw (Strozzi)--(Ridolfi);
    \draw (Strozzi)--(Castellani);
    \draw (Strozzi)--(Bischeri);
    \draw (Strozzi)--(Peruzzi);
    \draw (Tornabuoni)--(Guadagni);
    \draw (Tornabuoni)--(Ridolfi);
    \draw (Tornabuoni)--(Medici);
    \draw (Medici)--(Barbadori);
    \draw (Medici)--(Albizzi);
    \draw (Medici)--(Acciaiuoli);
    \draw (Medici)--(Salviati);
    \draw (Medici)--(Ridolfi);
    \draw (Albizzi)--(Guadagni);
    \draw (Albizzi)--(Ginori);
    \draw (Guadagni)--(Bischeri);
    \draw (Guadagni)--(Lamberteschi);
    \draw (Pazzi)--(Salviati);
    \draw (Bischeri)--(Peruzzi);
    \draw (Peruzzi)--(Castellani);
    \draw (Castellani)--(Barbadori);
}

\newcommand{\florentine}{
  \node (Acciaiuoli)   [v, label=$a$]at (0.41, 0.77) {};
  \node (Albizzi)      [v, label=$z$] at (0.20, 0.36) {};
  \node (Barbadori)    [v, label=$b$] at (0.62, 0.50) {};
  \node (Bischeri)     [v, label=below:$i$] at (0.55, 0.12) {};
  \node (Castellani)   [v, label=right:$c$] at (0.75, 0.34) {};
  \node (Ginori)       [v, label=left:$g$] at (0.00, 0.35) {};
  \node (Guadagni)     [v, label=below:$d$] at (0.32, 0.18) {};
  \node (Lamberteschi) [v, label=left:$l$] at (0.22, 0.00) {};
  \node (Medici)       [v, label=30:$m$] at (0.38, 0.54) {};
  \node (Pazzi)        [v, label=$p$] at (0.10, 0.75) {};
  \node (Peruzzi)      [v, label=right:$e$] at (0.73, 0.16) {};
  \node (Pucci)        [v, label=$u$] at (0.7, 0.7) {};
  \node (Ridolfi)      [v, label=below:$r$] at (0.50, 0.39) {};
  \node (Salviati)     [v, label=$v$] at (0.24, 0.65) {};
  \node (Strozzi)      [v, label=$s$] at (0.61, 0.28) {};
  \node (Tornabuoni)   [v, label=left:$t$] at (0.36, 0.36) {};
    \draw (Strozzi)--(Ridolfi);
    \draw (Strozzi)--(Castellani);
    \draw (Strozzi)--(Bischeri);
    \draw (Strozzi)--(Peruzzi);
    \draw (Tornabuoni)--(Guadagni);
    \draw (Tornabuoni)--(Ridolfi);
    \draw (Tornabuoni)--(Medici);
    \draw (Medici)--(Barbadori);
    \draw (Medici)--(Albizzi);
    \draw (Medici)--(Acciaiuoli);
    \draw (Medici)--(Salviati);
    \draw (Medici)--(Ridolfi);
    \draw (Albizzi)--(Guadagni);
    \draw (Albizzi)--(Ginori);
    \draw (Guadagni)--(Bischeri);
    \draw (Guadagni)--(Lamberteschi);
    \draw (Pazzi)--(Salviati);
    \draw (Bischeri)--(Peruzzi);
    \draw (Peruzzi)--(Castellani);
    \draw (Castellani)--(Barbadori);
}

\makeatletter
\def\thickhrule{\leavevmode \leaders \hrule height 1ex \hfill \kern \z@}
\def\position{\centering}
\renewcommand{\@makechapterhead}[1]{%
  {\parindent \z@ \position \reset@font
        {\huge \scshape  \thechapter }
        \par\nobreak
        \vspace*{10\p@}%
        {\huge \bfseries #1\par\nobreak}
    \vskip 40\p@
    \vskip 20\p@
  }}
\renewcommand\section{\@startsection{section}{1}{\z@}%
  {-3.5ex \@plus -1ex \@minus -.2ex}%
  {2.3ex \@plus.2ex}%
  {\normalfont\Large\bfseries\centering}}
\renewcommand\subsection{\@startsection{subsection}{2}{\z@}%
  {-3.5ex \@plus -1ex \@minus -.2ex}%
  {2.3ex \@plus.2ex \@minus -.2ex}%
  {\normalfont\bfseries\centering}}
\renewcommand{\@seccntformat}[1]{\csname the#1\endcsname. }
\renewcommand\paragraph{\@startsection{paragraph}{4}{\z@}%
  {3.25ex \@plus1ex \@minus.2ex}%
  {-1em}%
  {\normalfont\normalsize\bfseries}}
\makeatother

\usepackage{enumitem}
\newenvironment{algor}[3]{%
\bigskip
\noindent{\bf Algorithm #1} ({\it#2\/}) {\it #3}
\begin{description}\vspace{-.8ex}\setlength{\itemsep}{-2pt}\leftmargin=0pt}{%
\end{description}\medskip}
\setdescription{leftmargin=20pt}

\begin{document}
\setcounter{chapter}{12} 
\chapter{Graph colouring algorithms}

\vspace{-2cm}

\centerline{\sc Thore Husfeldt}

\bigskip
\noindent
\begin{minipage}{7cm}
\begin{tabbing}
1. \= Introduction\\
2. \> Greedy colouring\\
3. \> Local augmentation\\
4. \> Recursion\\
5. \> Subgraph expansion\\
6. \> Vector colouring\\
7. \> Reductions\\
References
\end{tabbing}
\end{minipage}

\bigskip
\begin{quote}\small \it
  This chapter presents an introduction to graph colouring algorithms.\footnote{
    Appears as Thore Husfeldt, \emph{Graph colouring algorithms},
    Chapter XIII of \emph{Topics in Chromatic Graph Theory}, L. W. Beineke and Robin J. Wilson (eds.),
    Encyclopedia of Mathematics and its Applications,
    Cambridge University Press, ISBN 978-1-107-03350-4, 2015, pp. 277--303.
  }
  The focus is on vertex-colouring algorithms that work for general
  classes of graphs with worst-case performance guarantees in a
  sequential model of computation.
  The presentation aims to demonstrate the breadth of available
  techniques and is organized by algorithmic paradigm.
\end{quote}

\section{Introduction}

A straightforward algorithm for finding a vertex-colouring of a graph
is to search systematically among all mappings from the set of
vertices to the set of colours, a technique often called
\emph{exhaustive} or \emph{brute force}:

\begin{algor}{X}{Exhaustive search}{Given an integer $q\geq 1$ and a
    graph $G$ with vertex set $V$, this algorithm finds a vertex-colouring using $q$ colours
    if one exists.}
\item[X1]
  [Main loop] For each mapping $\col\colon V\rightarrow
  \{1,2,\ldots,q\}$, do Step X2.
  \item[X2] [Check $\col$] If every edge $vw$ satisfies $\col(v)\neq
    \col(w)$, terminate with $f$ as the result. \qed
\end{algor}

This algorithm has few redeeming qualities, other than its being
correct.
We consider it here because it serves as an opportunity to make
explicit the framework in which we present more interesting
algorithms.

\subsection*{Model of computation}
If $G$ has $n$ vertices and $m$ edges, then the number of operations
used by Algorithm X can be asymptotically bounded by $O(q^n(n+m))$,
which we call the \emph{running time} of the algorithm.

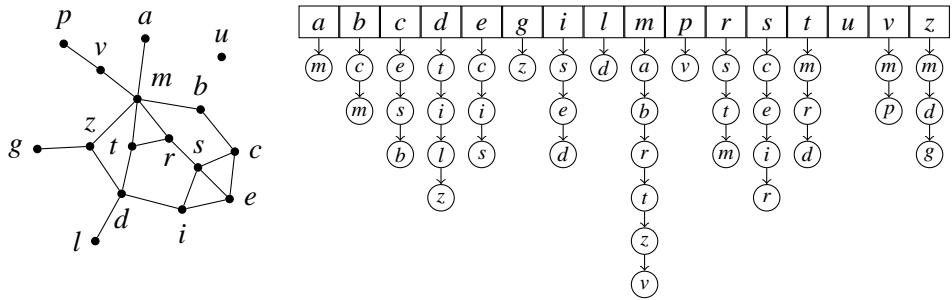
\begin{figure}[ht]
\centerline{
\raisebox{.5cm}{
\begin{tikzpicture}[scale=3.5]
\florentine
\end{tikzpicture}
}
\,
\begin{tikzpicture}[start chain,node distance=6pt and 0pt,
  every join/.style=->,
  list/.style={draw,on chain, circle, minimum width = 10pt,inner
    sep=0pt,join,font=\scriptsize\sf},
  array/.style={draw,on chain, rectangle, minimum width = 15pt,
    minimum height=12pt,
    inner sep=0pt,
    text height=1.7ex,text depth=.4ex,
    font=\small\sf}
]
  \node [array] {$a$};
  \begin{scope}[start branch=1 going below]
    \foreach \i in {m}
    \node [list] {$\i$} ;
  \end{scope}
  \node [array] {$b$};
  \begin{scope}[start branch= 2 going below]
    \foreach \i in {c,m}
    \node [list] {$\i$} ;
    \end{scope}
  \node [array] {$c$};
  \begin{scope}[start branch= 3 going below]
    \foreach \i in {e,s,b}
    \node [list] {$\i$} ;
  \end{scope}
  \node [array] {$d$};
  \begin{scope}[start branch= 4 going below]
    \foreach \i in {t,i,l,z}
    \node [list] {$\i$} ;
    \end{scope}
  \node [array] {$e$};
  \begin{scope}[start branch= 5 going below]
    \foreach \i in {c,i,s}
    \node [list] {$\i$} ;
    \end{scope}
  \node [array] {$g$};
  \begin{scope}[start branch= 6 going below]
    \foreach \i in {z}
    \node [list] {$\i$} ;
    \end{scope}
  \node [array] {$i$};
  \begin{scope}[start branch= 7 going below]
    \foreach \i in {s,e,d}
    \node [list] {$\i$} ;
    \end{scope}
  \node [array] {$l$};
  \begin{scope}[start branch= 8 going below]
    \foreach \i in {d}
    \node [list] {$\i$} ;
    \end{scope}
  \node [array] {$m$};
  \begin{scope}[start branch= 9 going below]
    \foreach \i in {a,b,r,t,z,v}
    \node [list] {$\i$} ;
    \end{scope}
  \node [array] {$p$};
  \begin{scope}[start branch= 10 going below]
    \foreach \i in {v}
    \node [list] {$\i$} ;
    \end{scope}
  \node [array] {$r$};
  \begin{scope}[start branch= 11 going below]
    \foreach \i in {s,t,m}
    \node [list] {$\i$};
    \end{scope}
  \node [array] {$s$};
  \begin{scope}[start branch= 12 going below]
    \foreach \i in {c,e,i,r}
    \node [list] {$\i$} ;
    \end{scope}
  \node [array] {$t$};
  \begin{scope}[start branch= 13 going below]
    \foreach \i in {m,r,d}
    \node [list] {$\i$} ;
    \end{scope}
  \node [array] {$u$};
  \begin{scope}[start branch= 14 going below]
    \foreach \i in {}
    \node [list] {\i} ;
    \end{scope}
  \node [array] {$v$};
  \begin{scope}[start branch= 15 going below]
    \foreach \i in {m,p}
    \node [list] {$\i$} ;
    \end{scope}
  \node [array] {$z$};
  \begin{scope}[start branch= 16 going below]
    \foreach \i in {m,d,g}
    \node [list] {$\i$} ;
    \end{scope}
\end{tikzpicture}
}
\caption{\label{fig: cap}A graph and its representation as an array of
  sequences}
\end{figure}

To make such a claim, we tacitly assume a computational model that
includes primitive operations, such as iterating over all mappings
from one finite set $A$ to another finite set $B$ in time
$O(|B|^{|A|})$ (Step X1), or iterating over all edges in time $O(n+m)$
(Step X2).
For instance, we assume that the input graph is represented by an
array of sequences indexed by vertices; the sequence stored at vertex
$v$ contains the neighouring vertices $N(v)$, see Fig.~\ref{fig: cap}.
This representation allows us to iterate over the neighbours of a
vertex in time $O(\deg v)$.
(An alternative representation, such as an incidence or adjacency
matrix, would not allow this.)
Note that detecting whether two graphs are isomorphic is \emph{not} a
primitive operation.
The convention of expressing computational resources using asymptotic
notation is consistent with our somewhat cavalier attitude towards the
details of our computational model.
Our assumptions are consistent with the behaviour of a modern computer
in a high-level programming language.
Nevertheless, we will explain our algorithms in plain English.

\subsection*{Worst-case asymptotic analysis}

Note that we could have fixed the colouring of a specific vertex $v$
as $\col(v)=0$, reducing Algorithm X's running time to
$O(q^{n-1}(n+m))$.
A moment's thought shows that this reasoning can then be extended to
cliques of size $r\geq 1$: search through all $\binom{n}{r}$ induced
subgraphs until a clique of size $r$ is found, arbitrarily map these
vertices to $\{1,2,\ldots,r\}$ and then let Algorithm X colour the
remaining vertices.
This reduces the running time to $O(q^{n-\omega(G)} n^{\omega(G)}(n+
m))$, where $\omega(G)$ is the clique size.
This may be quite useful for some graphs.
Another observation is that in the best case, the running time is
$O(n+m)$.
However, we will normally not pursue this kind of argument.
Instead, we are maximally pessimistic about the input and the
algorithm's underspecified choices.
In other words, we understand running times as worst-case performance
guarantees, rather than `typical' running times or average running
times over some distribution.

Sometimes we may even say that Algorithm X requires time
$q^n\poly(n)$, where we leave the polynomial factor unspecified in
order to signal the perfunctory attention we extend to these issues.

\subsection*{Overview and notation}

Straightforward variants of Algorithm X can be used to solve some
other graph colouring problems.
For instance, to find a list-colouring, we restrict the range of
values for each $\col(v)$ to a given list; to find an edge-colouring,
we iterate over all mappings $\col\colon
E\rightarrow\{1,2,\ldots,q\}$.

Another modification is to count the number of colourings instead of
finding just one.
These extensions provide baseline algorithms for list-colouring,
edge-colouring, the chromatic polynomial, the chromatic index, and so
forth.
However, for purposes of exposition, we present algorithms in their
\emph{least} general form, emphasizing the algorithmic idea rather
than its (sometimes quite pedestrian) generalizations.
The algorithms are organized by algorithmic technique rather than
problem type, graph class, optimality criterion, or computational
complexity.
These sections are largely independent and can be read in any order,
except perhaps for Algorithm G in Section~\ref{sec: greedy}.
The final section takes a step back and relates
the various colouring problems to each other.

\section{Greedy colouring}
\label{sec: greedy}
The following algorithm, sometimes called the \emph{greedy} or
\emph{sequential} algorithm, considers the vertices one by one and uses the
first available colour.

\begin{algor}{G}{Greedy vertex-colouring}{ Given a graph $G$ with
    maximum degree $\Delta$ and an ordering $v_1,v_2,\ldots,v_n$ of
    its vertices, this algorithm finds a vertex-colouring with
    $\max_{i} |\{\, j< i \colon v_jv_i \in E\,\}|+1 \leq \Delta + 1$
    colours.}
  \item[G1] [Initialize] Set $i=0$.
  \item[G2] [Next vertex]
    Increment $i$.
    If $i=n+1$, terminate with $\col$ as the result.
  \item[G3]
    [Find the colours $N(v_i)$] Compute the set
    $C=\bigcup_{j<i} \col(v_j)$ of colours already assigned to the
    neighbours of $v_i$.
    \item[G4] [Assign the smallest available colour to $v_i$]
      For increasing $c=1,2,\dots$, check whether $c\in C$. If not, set
      $\col(v_i)= c$ and return to Step G2. \qed
\end{algor}

For the number of colours, it is clear that in Step G4, the value of
$c$ is at most $|C|$, which is bounded by the number of neighbours of $v_i$
among $v_1,v_2, \ldots,v_{i-1}$.
In particular, Algorithm G establishes that $\chi(G)\leq \Delta(G)+1$.

For the running time, note that both Steps G3 and G4 take at most
$O(1+\deg v_i)$ operations.
Summing over all $i$, the total time spent in Steps G3 and G4 is
asymptotically bounded by $n+(\deg v_1+\deg v_2+\cdots+\deg v_n) = n+2m$.
Thus, Algorithm G takes time $O(n+m)$.

\begin{figure}[ht]
\centerline{    \begin{tikzpicture}[xscale=.733]
      \node (0) [v,label=below:$v_2$] at (0,0) {};
      \node (1) [v,label=above:$v_1$] at (0,1) {};
      \node (2) [v,label=below:$v_4$] at (1,0) {};
      \node (3) [v,label=above:$v_3$] at (1,1) {};
      \node (4) [v,label=below:$v_6$] at (2,0) {};
      \node (5) [v,label=above:$v_5$] at (2,1) {};
      \node  (6) at (2.75,0) {$\cdots$};
      \node  (7) at (2.75,1) {$\cdots$};
      \node (8) [v,label=below:$v_n$] at (4,0) {};
      \node (9) [v,label=above:$v_{n-1}$] at (4,1) {};
      \draw (0)--(3);   \draw (0)--(9);
      \draw (2)--(1);  \draw (2)--(9);
      \draw (4)--(1); \draw (4)--(3); \draw (4)--(9);
      \draw (5)--(0); \draw (5)--(2); \draw (5)--(8);
      \draw (8)--(1); \draw (8)--(3);
    \end{tikzpicture}}
\caption{\label{fig: crown graph} The crown graph}
\end{figure}
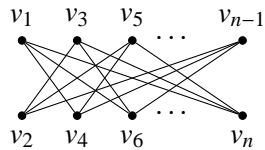

\paragraph{Optimal ordering}
The size of the colouring computed by Algorithm G depends heavily on
the vertex ordering.
Its worst-case behaviour is poor.
For instance, it spends $\frac{1}{2}n$ colours on the 2-colourable \emph{crown
  graph} shown in Fig.~\ref{fig: crown graph}.

On the other hand, for every graph there exists an ordering for which
Algorithm G uses an optimal number of colours; indeed, any ordering
that satisfies $\col(v_i)\leq \col(v_{i+1})$ for an optimal colouring
$\col$ has this property.
Since there are $n!$ different orderings, this observation is
algorithmically quite useless.
An ordering is \emph{perfect} for a graph if, for every induced
subgraph, Algorithm G results in an optimal colouring; triangulated
graphs and comparability graphs always admit such an ordering, as
shown by Chvátal~\cite{chvatal}.

\subsection*{Randomness}
Algorithm G performs quite well on random graphs, whatever the vertex
ordering.
For almost all $n$-vertex graphs, it uses $n/(\log n- 3\log\log n)$
colours, which is roughly twice the optimum value (see \cite{GMcD}).

This suggests the following randomized algorithm. For a graph $G$,
choose a vertex ordering at random and then execute Algorithm G.
For many problems, it is a sound algorithmic design strategy
to trade good average-case behaviour for good (expected) worst-case
behaviour in this way.
However, for Algorithm G the result is quite poor:
for every
$\epsilon>0$ there exist graphs with chromatic number $n^\epsilon$
for which the randomized algorithm uses $\Omega(n/\log n)$ colours
with high probability, as shown by Ku\v{c}era \cite{Kucera}.

\subsection*{Other orderings}
In the \emph{largest-first} vertex-degree ordering introduced by Welsh and
Powell \cite{WP}, the vertices are ordered such that $\deg v_1 \geq
\deg v_2 \geq\cdots\geq \deg v_n$.
This establishes the bound $\chi(G) \leq 1+\max_i\min \{\deg
v_i,i-1\}$, which is sometimes better than $1+\Delta$, such as in
Fig.~\ref{fig: C5vee1}.

\begin{figure}[ht]
\centerline{
    \begin{tikzpicture}[scale=.75]
      \node (1) [v,label=below:$v_1$] at (0,0) {};
      \node (2) [v,label=90+0*72:$v_2$] at (90+0*72:1) {};
      \node (3) [v,label=90+1*72:$v_3$] at (90+1*72:1) {};
      \node (4) [v,label=90+2*72:$v_4$] at (90+2*72:1) {};
      \node (5) [v,label=90+3*72:$v_5$] at (90+3*72:1) {};
      \node (6) [v,label=90+4*72:$v_6$] at (90+4*72:1) {};
      \draw (2)--(3)--(4)--(5)--(6)--(2);
      \foreach \i in {2,3,4,5,6}
      \draw (1)--(\i);
    \end{tikzpicture}}
\caption{\label{fig: C5vee1}\ }
\end{figure}

Closely related in spirit is Matula's \emph{smallest-last} ordering
\cite{M}, given as follows: choose as the last vertex $v_n$ a vertex
of minimum degree in $G$, and proceed recursively with $G - v_n$, see Fig.~\ref{fig: orderings}.
With this ordering, the size of the resulting colouring is be bounded
by the Szekeres--Wilf bound \cite{SW}, \[\chi(G)\leq
\operatorname{dgn}(G)+1\,,\] where the \emph{degeneracy}
$\operatorname{dgn}(G)$ is the maximum over all subgraphs $H$ of $G$
of the minimum degree $\delta(H)$.
This ordering optimally colours crown graphs and many other classes of
graphs, and uses six colours on any planar graph.

\begin{figure}[ht]
\centerline{\begin{tikzpicture}[scale=.5]
\node [anchor = east] at (-.5,0) {largest-first:};
\node (0) [v, label = $v_6$] at (0,0){};
\node (1) [v, label = $v_1$] at (1,0){};
\node (2) [v, label = $v_4$] at (2,0){};
\node (3) [v, label = $v_3$] at (3,0){};
\node (4) [v, label = $v_2$] at (4,0){};
\node (5) [v, label = $v_5$] at (5,0){};
\draw (0)--(1)--(2)--(3)--(4)--(5);
\end{tikzpicture}
\qquad
\begin{tikzpicture}[scale=.5]
\node [anchor = east] at (-.5,0) {smallest-last:};
\node (0) [v, label = $v_6$] at (0,0){};
\node (1) [v, label = $v_4$] at (1,0){};
\node (2) [v, label = $v_2$] at (2,0){};
\node (3) [v, label = $v_1$] at (3,0){};
\node (4) [v, label = $v_3$] at (4,0){};
\node (5) [v, label = $v_5$] at (5,0){};
\draw (0)--(1)--(2)--(3)--(4)--(5);
\end{tikzpicture}
}
\caption{\label{fig: orderings}\ }
\end{figure}

Other orderings are dynamic in the sense that the ordering is
determined during the execution of the algorithm, rather than in advance.
For example, Brélaz \cite{Brel} suggests choosing the next vertex from among those adjacent
to the largest number of different colours.
Many other orderings have been investigated (see the surveys of Kosowski and Manuszewski~\cite{KM} and Maffray~\cite{maffray}).
Many of them perform quite well on instances that one may encounter
`in practice', but attempts at formalizing what this means are
quixotic.

\subsection*{2-colourable graphs}

Of particular interest are those vertex orderings in which every
vertex $v_i$ is adjacent to some vertex $v_j$ with $j<i$.
Such orderings can be computed in time $O(m+n)$ using basic graph-traversal algorithms.
This algorithm is sufficiently important to be made explicit.

\begin{algor}{B}{Bipartition}{
  Given a connected graph $G$, this algorithm finds a 2-colouring if one exists.
  Otherwise, it outputs an odd cycle.}
\item[B1]
  [Initialize]
  Let $\col(v_1)=1$ and let $Q$ (the `queue') be an empty sequence.
  For each neighbour $w$ of $v_1$, set $p(w)=v_1$ (the `parent' of
  $w$) and add $w$ to $Q$.
  \item[B2]
    [Next vertex] If $Q$ is empty, go to  Step B3.
    Otherwise, remove the first vertex $v$ from $Q$ and
    set $\col(v)$ to the colour not already assigned to $p(v)$.
    For each neighbour $w$ of $v$, if $w$ is not yet coloured and
    does not belong to $Q$, then set $p(w)=v$ and add $w$ to the end of $Q$.
    Repeat Step B2.
  \item[B3]
    [Verify 2-colouring] Iterate over all edges to verify that $\col(v)\neq
    \col(w)$ for every edge $vw$.
    If so, terminate with $f$ as the result.
  \item[B4]
    [Construct odd cycle]
    Let $vw$ be an edge with $\col(v)=\col(w)$ and
    let $u$ be the nearest common ancestor of $v$ and $w$ in the
    tree defined by $p$.
    Output the path $w,p(w),p(p(w)),\ldots,u$, followed by the reversal of the
    path $v,p(v),p(p(v)),\allowbreak\ldots,\allowbreak u$, followed by the edge $vw$.
    \qed
\end{algor}

\begin{figure}[ht]
\centerline{\begin{tikzpicture}[scale=.13,  every node/.style={v}, gray]
\women
\node [draw,black,very thick,fill=white,inner sep=2pt] at (E9) {};
\end{tikzpicture}\qquad
\begin{tikzpicture}[scale=.13,  every node/.style={v}, gray]
\women
\begin{scope}[very thick,every node/.style={inner sep =2pt,circle}]
\draw [black] (Dorothy Murchison)--(E9);
\draw [black] (Flora Price)--(E9);
\draw [black] (Katherina Rogers)--(E9);
\draw [black] (Nora Fayette)--(E9);
\draw [black] (Olivia Carleton)--(E9);
\draw [black] (Pearl Oglethorpe)--(E9);
\draw [black] (Sylvia Avondale)--(E9);
\draw [black] (Theresa Anderson)--(E9);
\draw [black] (Verne Sanderson)--(E9);
\node [draw,black,fill=lightgray] at (Dorothy Murchison) {};
\node [draw,black,fill=lightgray] at (Flora Price) {};
\node [draw,black,fill=lightgray] at (Katherina Rogers) {};
\node [draw,black,fill=lightgray] at (Nora Fayette) {};
\node [draw,black,fill=lightgray] at (Olivia Carleton) {};
\node [draw,black,fill=lightgray] at (Pearl Oglethorpe) {};
\node [draw,black,fill=lightgray] at (Sylvia Avondale) {};
\node [draw,black,fill=lightgray] at (Theresa Anderson) {};
\node [draw,black,fill=lightgray] at (Verne Sanderson) {};
\node [draw,black,fill=white] at (E9) {};
\end{scope}
\end{tikzpicture}
\qquad
\begin{tikzpicture}[scale=.13,  every node/.style={v}, gray]
\women
\begin{scope}[very thick, every node/.style={inner sep = 2pt,circle}]
\draw [black] (Dorothy Murchison)--(E9);
\draw [black] (E10)--(Katherina Rogers);
\draw [black] (E11)--(Flora Price);
\draw [black] (E2)--(Theresa Anderson);
\draw [black] (E3)--(Theresa Anderson);
\draw [black] (E4)--(Theresa Anderson);
\draw [black] (E5)--(Theresa Anderson);
\draw [black] (E6)--(Theresa Anderson);
\draw [black] (E7)--(Theresa Anderson);
\draw [black] (E8)--(Theresa Anderson);
\draw [black] (Evelyn Jefferson)--(E9);
\draw [black] (Flora Price)--(E9);
\draw [black] (Katherina Rogers)--(E9);
\draw [black] (Myra Liddel)--(E9);
\draw [black] (Nora Fayette)--(E9);
\draw [black] (Olivia Carleton)--(E9);
\draw [black] (Pearl Oglethorpe)--(E9);
\draw [black] (Ruth DeSand)--(E9);
\draw [black] (Sylvia Avondale)--(E9);
\draw [black] (Theresa Anderson)--(E9);
\draw [black] (Verne Sanderson)--(E9);
\node [draw,black,fill=lightgray] at (Dorothy Murchison) {};
\node [draw,black,fill=lightgray] at (Evelyn Jefferson) {};
\node [draw,black,fill=lightgray] at (Flora Price) {};
\node [draw,black,fill=lightgray] at (Katherina Rogers) {};
\node [draw,black,fill=lightgray] at (Myra Liddel) {};
\node [draw,black,fill=lightgray] at (Nora Fayette) {};
\node [draw,black,fill=lightgray] at (Olivia Carleton) {};
\node [draw,black,fill=lightgray] at (Pearl Oglethorpe) {};
\node [draw,black,fill=lightgray] at (Ruth DeSand) {};
\node [draw,black,fill=lightgray] at (Sylvia Avondale) {};
\node [draw,black,fill=lightgray] at (Theresa Anderson) {};
\node [draw,black,fill=lightgray] at (Verne Sanderson) {};
\node [draw,black,fill=white] at (E10) {};
\node [draw,black,fill=white] at (E11) {};
\node [draw,black,fill=white] at (E2) {};
\node [draw,black,fill=white] at (E3) {};
\node [draw,black,fill=white] at (E4) {};
\node [draw,black,fill=white] at (E5) {};
\node [draw,black,fill=white] at (E6) {};
\node [draw,black,fill=white] at (E7) {};
\node [draw,black,fill=white] at (E8) {};
\node [draw,black,fill=white] at (E9) {};
\end{scope}
\end{tikzpicture}
\qquad
\begin{tikzpicture}[scale=.13,  every node/.style={v}, gray]
\women
\begin{scope}[very thick, every node/.style={inner sep = 2pt,circle}]
\draw [black] (Brenda Rogers)--(E8);
\draw [black] (Charlotte McDowd)--(E5);
\draw [black] (Dorothy Murchison)--(E9);
\draw [black] (E1)--(Evelyn Jefferson);
\draw [black] (E10)--(Katherina Rogers);
\draw [black] (E11)--(Flora Price);
\draw [black] (E12)--(Katherina Rogers);
\draw [black] (E13)--(Katherina Rogers);
\draw [black] (E14)--(Katherina Rogers);
\draw [black] (E2)--(Theresa Anderson);
\draw [black] (E3)--(Theresa Anderson);
\draw [black] (E4)--(Theresa Anderson);
\draw [black] (E5)--(Theresa Anderson);
\draw [black] (E6)--(Theresa Anderson);
\draw [black] (E7)--(Theresa Anderson);
\draw [black] (E8)--(Theresa Anderson);
\draw [black] (Eleanor Nye)--(E8);
\draw [black] (Evelyn Jefferson)--(E9);
\draw [black] (Flora Price)--(E9);
\draw [black] (Frances Anderson)--(E8);
\draw [black] (Helen Lloyd)--(E8);
\draw [black] (Katherina Rogers)--(E9);
\draw [black] (Laura Mandeville)--(E8);
\draw [black] (Myra Liddel)--(E9);
\draw [black] (Nora Fayette)--(E9);
\draw [black] (Olivia Carleton)--(E9);
\draw [black] (Pearl Oglethorpe)--(E9);
\draw [black] (Ruth DeSand)--(E9);
\draw [black] (Sylvia Avondale)--(E9);
\draw [black] (Theresa Anderson)--(E9);
\draw [black] (Verne Sanderson)--(E9);
\node [draw,black,fill=lightgray] at (Brenda Rogers) {};
\node [draw,black,fill=lightgray] at (Charlotte McDowd) {};
\node [draw,black,fill=lightgray] at (Dorothy Murchison) {};
\node [draw,black,fill=lightgray] at (Eleanor Nye) {};
\node [draw,black,fill=lightgray] at (Evelyn Jefferson) {};
\node [draw,black,fill=lightgray] at (Flora Price) {};
\node [draw,black,fill=lightgray] at (Frances Anderson) {};
\node [draw,black,fill=lightgray] at (Helen Lloyd) {};
\node [draw,black,fill=lightgray] at (Katherina Rogers) {};
\node [draw,black,fill=lightgray] at (Laura Mandeville) {};
\node [draw,black,fill=lightgray] at (Myra Liddel) {};
\node [draw,black,fill=lightgray] at (Nora Fayette) {};
\node [draw,black,fill=lightgray] at (Olivia Carleton) {};
\node [draw,black,fill=lightgray] at (Pearl Oglethorpe) {};
\node [draw,black,fill=lightgray] at (Ruth DeSand) {};
\node [draw,black,fill=lightgray] at (Sylvia Avondale) {};
\node [draw,black,fill=lightgray] at (Theresa Anderson) {};
\node [draw,black,fill=lightgray] at (Verne Sanderson) {};
\node [draw,black,fill=white] at (E1) {};
\node [draw,black,fill=white] at (E10) {};
\node [draw,black,fill=white] at (E11) {};
\node [draw,black,fill=white] at (E12) {};
\node [draw,black,fill=white] at (E13) {};
\node [draw,black,fill=white] at (E14) {};
\node [draw,black,fill=white] at (E2) {};
\node [draw,black,fill=white] at (E3) {};
\node [draw,black,fill=white] at (E4) {};
\node [draw,black,fill=white] at (E5) {};
\node [draw,black,fill=white] at (E6) {};
\node [draw,black,fill=white] at (E7) {};
\node [draw,black,fill=white] at (E8) {};
\node [draw,black,fill=white] at (E9) {};
\end{scope}
\end{tikzpicture}}
\caption{\label{fig: Algorithm B}Execution of Algorithm B}
\end{figure}
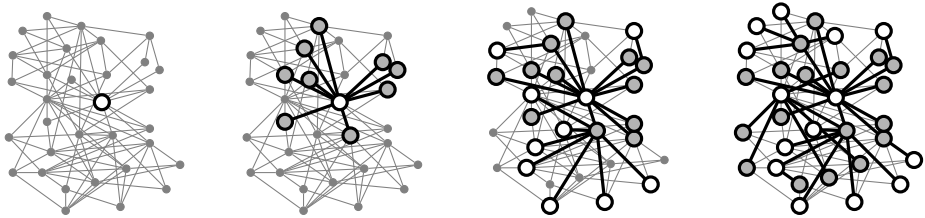

Fig.~\ref{fig: Algorithm B} shows an execution of Algorithm B finding
a 2-colouring.

Algorithm B is an example of a `certifying' algorithm: an algorithm
that produces a witness to certify its correctness, in this case an
odd cycle if the graph is not 2-colourable.
To see that the cycle constructed in Step~B4 has odd length, note that on the
two paths $w,p(w),p(p(w)),\ldots, u$ and $v,p(v),p(p(v)),\ldots,u$, each vertex
has a different colour from its predecessor.
Since the respective endpoints of both paths have the same colour,
they must contain the same number of edges modulo 2.
In particular, their total length is even.
With the additional edge $vw$, the length of the resulting cycle is odd.

\begin{figure}[ht]
\centerline{\begin{tikzpicture}[scale=.13,  every node/.style={v},gray]
\women
\begin{scope}[every node/.style={inner sep = 2pt, circle}, very thick]
\node [draw,black,fill=white] at (E9) {};
\end{scope}
\end{tikzpicture}
\qquad
\begin{tikzpicture}[scale=.13,  every node/.style={v},gray]
\women
\begin{scope}[every node/.style={inner sep = 2pt, circle}, very thick]
\draw [black] (Theresa Anderson)--(E9);
\draw [black] (Pearl Oglethorpe)--(E8);
\draw [black] (Nora Fayette)--(E6);
\draw [black] (E10)--(Helen Lloyd);
\draw [black] (Olivia Carleton)--(E11);
\draw [black] (E11)--(Nora Fayette);
\draw [black] (E8)--(Theresa Anderson);
\draw [black] (E6)--(Pearl Oglethorpe);
\draw [black] (Helen Lloyd)--(E11);
\node [draw,black,fill=lightgray] at (Theresa Anderson) {};
\node [draw,black,fill=lightgray] at (Pearl Oglethorpe) {};
\node [draw,black,fill=lightgray] at (Nora Fayette) {};
\node [draw,black,fill=white] at (E10) {};
\node [draw,black,fill=lightgray] at (Olivia Carleton) {};
\node [draw,black,fill=white] at (E11) {};
\node [draw,black,fill=white] at (E9) {};
\node [draw,black,fill=white] at (E8) {};
\node [draw,black,fill=white] at (E6) {};
\node [draw,black,fill=lightgray] at (Helen Lloyd) {};
\end{scope}
\end{tikzpicture}
\qquad
\begin{tikzpicture}[scale=.13,  every node/.style={v},gray]
\women
\begin{scope}[every node/.style={inner sep = 2pt, circle}, very thick]
\draw [black] (Nora Fayette)--(E6);
\draw [black] (E10)--(Helen Lloyd);
\draw [black] (E13)--(Sylvia Avondale);
\draw [black] (E12)--(Katherina Rogers);
\draw [black] (Frances Anderson)--(E5);
\draw [black] (Helen Lloyd)--(E11);
\draw [black] (Olivia Carleton)--(E11);
\draw [black] (Eleanor Nye)--(E5);
\draw [black] (E11)--(Nora Fayette);
\draw [black] (E8)--(Theresa Anderson);
\draw [black] (E5)--(Brenda Rogers);
\draw [black] (E7)--(Verne Sanderson);
\draw [black] (E6)--(Pearl Oglethorpe);
\draw [black] (Brenda Rogers)--(E7);
\draw [black] (E3)--(Frances Anderson);
\draw [black] (Katherina Rogers)--(E13);
\draw [black] (Laura Mandeville)--(E3);
\draw [black] (Pearl Oglethorpe)--(E8);
\draw [black] (Sylvia Avondale)--(E10);
\draw [black] (Theresa Anderson)--(E9);
\draw [black] (Verne Sanderson)--(E12);
\node [draw,black,fill=lightgray] at (Nora Fayette) {};
\node [draw,black,fill=white] at (E10) {};
\node [draw,black,fill=white] at (E13) {};
\node [draw,black,fill=white] at (E12) {};
\node [draw,black,fill=lightgray] at (Frances Anderson) {};
\node [draw,black,fill=lightgray] at (Helen Lloyd) {};
\node [draw,black,fill=lightgray] at (Olivia Carleton) {};
\node [draw,black,fill=lightgray] at (Eleanor Nye) {};
\node [draw,black,fill=white] at (E11) {};
\node [draw,black,fill=white] at (E9) {};
\node [draw,black,fill=white] at (E8) {};
\node [draw,black,fill=white] at (E5) {};
\node [draw,black,fill=white] at (E7) {};
\node [draw,black,fill=white] at (E6) {};
\node [draw,black,fill=lightgray] at (Verne Sanderson) {};
\node [draw,black,fill=lightgray] at (Brenda Rogers) {};
\node [draw,black,fill=lightgray] at (Katherina Rogers) {};
\node [draw,black,fill=lightgray] at (Laura Mandeville) {};
\node [draw,black,fill=lightgray] at (Pearl Oglethorpe) {};
\node [draw,black,fill=lightgray] at (Sylvia Avondale) {};
\node [draw,black,fill=lightgray] at (Theresa Anderson) {};
\node [draw,black,fill=white] at (E3) {};
\end{scope}
\end{tikzpicture}
\qquad
\begin{tikzpicture}[scale=.13,  every node/.style={v},gray]
\women
\begin{scope}[every node/.style={inner sep = 2pt,circle}, very thick]
\draw [black] (Brenda Rogers)--(E7);
\draw [black] (Charlotte McDowd)--(E4);
\draw [black] (Dorothy Murchison)--(E8);
\draw [black] (E1)--(Laura Mandeville);
\draw [black] (E10)--(Helen Lloyd);
\draw [black] (E11)--(Nora Fayette);
\draw [black] (E12)--(Katherina Rogers);
\draw [black] (E13)--(Sylvia Avondale);
\draw [black] (E14)--(Katherina Rogers);
\draw [black] (E2)--(Evelyn Jefferson);
\draw [black] (E3)--(Frances Anderson);
\draw [black] (E4)--(Evelyn Jefferson);
\draw [black] (E5)--(Brenda Rogers);
\draw [black] (E6)--(Pearl Oglethorpe);
\draw [black] (E7)--(Verne Sanderson);
\draw [black] (E8)--(Theresa Anderson);
\draw [black] (Eleanor Nye)--(E5);
\draw [black] (Evelyn Jefferson)--(E1);
\draw [black] (Flora Price)--(E11);
\draw [black] (Frances Anderson)--(E5);
\draw [black] (Helen Lloyd)--(E11);
\draw [black] (Katherina Rogers)--(E13);
\draw [black] (Laura Mandeville)--(E3);
\draw [black] (Myra Liddel)--(E12);
\draw [black] (Nora Fayette)--(E6);
\draw [black] (Olivia Carleton)--(E11);
\draw [black] (Pearl Oglethorpe)--(E8);
\draw [black] (Ruth DeSand)--(E5);
\draw [black] (Sylvia Avondale)--(E10);
\draw [black] (Theresa Anderson)--(E9);
\draw [black] (Verne Sanderson)--(E12);
\node [draw,black,fill=lightgray] at (Brenda Rogers) {};
\node [draw,black,fill=lightgray] at (Charlotte McDowd) {};
\node [draw,black,fill=lightgray] at (Dorothy Murchison) {};
\node [draw,black,fill=lightgray] at (Eleanor Nye) {};
\node [draw,black,fill=lightgray] at (Evelyn Jefferson) {};
\node [draw,black,fill=lightgray] at (Flora Price) {};
\node [draw,black,fill=lightgray] at (Frances Anderson) {};
\node [draw,black,fill=lightgray] at (Helen Lloyd) {};
\node [draw,black,fill=lightgray] at (Katherina Rogers) {};
\node [draw,black,fill=lightgray] at (Laura Mandeville) {};
\node [draw,black,fill=lightgray] at (Myra Liddel) {};
\node [draw,black,fill=lightgray] at (Nora Fayette) {};
\node [draw,black,fill=lightgray] at (Olivia Carleton) {};
\node [draw,black,fill=lightgray] at (Pearl Oglethorpe) {};
\node [draw,black,fill=lightgray] at (Ruth DeSand) {};
\node [draw,black,fill=lightgray] at (Sylvia Avondale) {};
\node [draw,black,fill=lightgray] at (Theresa Anderson) {};
\node [draw,black,fill=lightgray] at (Verne Sanderson) {};
\node [draw,black,fill=white] at (E1) {};
\node [draw,black,fill=white] at (E10) {};
\node [draw,black,fill=white] at (E11) {};
\node [draw,black,fill=white] at (E12) {};
\node [draw,black,fill=white] at (E13) {};
\node [draw,black,fill=white] at (E14) {};
\node [draw,black,fill=white] at (E2) {};
\node [draw,black,fill=white] at (E3) {};
\node [draw,black,fill=white] at (E4) {};
\node [draw,black,fill=white] at (E5) {};
\node [draw,black,fill=white] at (E6) {};
\node [draw,black,fill=white] at (E7) {};
\node [draw,black,fill=white] at (E8) {};
\node [draw,black,fill=white] at (E9) {};
\end{scope}
\end{tikzpicture}}
\caption{\label{fig: dfs}Execution of Algorithm B using depth-first search}
\end{figure}
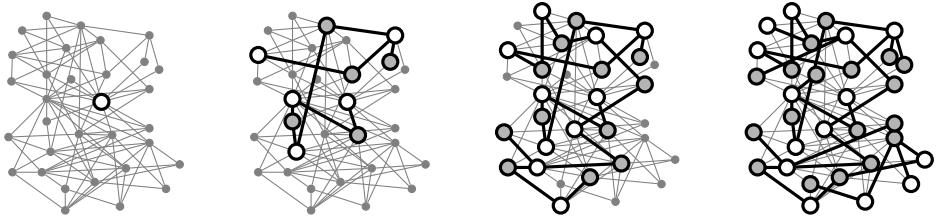

The order in which the vertices are considered by Algorithm B depends on
the first-in first-out behaviour of the queue $Q$.
The resulting ordering is called \emph{breadth-first}.
An important variant uses a last-in first-out `stack' instead of
a queue; the resulting ordering is called
\emph{depth-first}.
Fig.~\ref{fig: dfs} shows the resulting behaviour
on the graph from Fig.~\ref{fig: Algorithm B}.

\bigskip

Algorithm B works also for the list-colouring problem, provided that
for each vertex $v$, the available list of colours $L(v)$ has size at
most 2.
This observation leads to a simple, randomized, exponential-time
algorithm for 3-colouring due to Beigel and Eppstein \cite{BE}.

\begin{algor}{P}{Palette restriction}{Given a graph, this algorithm finds a 3-colouring if one exists.}
\item[P1] [Forbid one colour at each vertex] For each vertex $v$, select a list $L(v)$ of colours
  available at $v$ uniformly and independently at random from the three
  lists $\{1,2\}$, $\{2,3\}$, and $\{1,3\}$.
\item[P2] [Attempt 2-colouring] Try to solve the list-colouring instance given by $L$ using
  Algorithm B, setting $f(v_1)=\min L(v_1)$ in Step B1.
  If successful, terminate with the resulting colouring.
  Otherwise, return to Step P1. \qed
\end{algor}

To analyse the running time, consider a 3-colouring $f$.
For each vertex $v$, colour $f(v)$ belongs to $L(v)$ with probablity
$\frac{2}{3}$.
Thus, with probability at least $(\frac{2}{3})^n$, the list colouring instance
constructed in step P1 has a solution.
It follows that the expected number of repetitions is $(\frac{3}{2})^n$, each of which
takes polynomial time.

\subsection*{Wigderson's algorithm}

Algorithms~B and G appear together in  Wigderson's algorithm~\cite{W}:

\begin{algor}{W}{Wigderson's algorithm}
{  Given a $3$-chromatic graph $G$, this algorithm finds a vertex-colouring with
  $O(\surd n)$ colours.}
\item[W1]
  [Initialize] Let $c=1$.
  \item[W2]
    [$\Delta(G)\geq \lceil\surd n\rceil$] Consider a vertex $v$ in $G$ with
    $\deg v\geq \lceil \surd n\rceil$; if no such vertex exists,
    go to Step W3.
    Use Algorithm B to 2-colour the neighbourhood $G[N(v)]$ with
    colours $c$ and $c+1$.
    Remove $N(v)$ from $G$ and increase $c$ by $\chi(G[N(v)])$.
    Repeat Step W2.
  \item[W3] [$\Delta(G)<\lceil\surd n\rceil$] Use Algorithm G to colour the
    remaining vertices with the colours $c,c+1,\ldots, c+\lceil \surd
    n\rceil$. \qed
\end{algor}

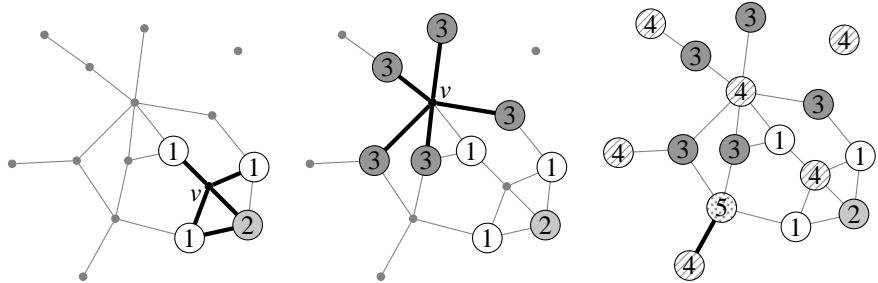
\begin{figure}[ht]
  \tikzset{every picture/.style={scale=3.3,font=\small}}
  \tikzset{every node/.style={draw,circle, fill, inner sep=0pt}}
\quad
  \begin{tikzpicture}[scale=1.3,gray]
   \florentineee
   \begin{scope}[black,
   every node/.style={draw=black, circle, inner sep=1pt}]
   \draw [ultra thick] (Strozzi)--(Castellani);
   \draw [ultra thick](Strozzi)--(Peruzzi);
   \draw [ultra thick](Strozzi)--(Bischeri);
   \draw [ultra thick](Strozzi)--(Ridolfi);
   \draw [ultra thick](Bischeri)--(Peruzzi);
   \node[v,label=185:\small $v$] at (Strozzi) {};
   \node[fill=col1] at (Castellani) {1};
   \node[fill=col2] at (Peruzzi) {2};
   \node[fill=col1] at (Bischeri) {1};
   \node[fill=col1] at (Ridolfi) {1};
   \end{scope}
  \end{tikzpicture}
\quad
  \begin{tikzpicture}[scale=1.3,gray]
   \florentineee
    \begin{scope}[ black,
   every node/.style={draw=black, circle, inner sep=1pt}]
   \draw [ultra thick](Medici)--(Salviati);
   \draw [ultra thick](Medici)--(Albizzi);
   \draw [ultra thick](Medici)--(Acciaiuoli);
   \draw [ultra thick](Medici)--(Tornabuoni);
   \draw [ultra thick](Medici)--(Barbadori);
   \node [fill=col1] at (Castellani) {1};
   \node [fill=col2] at (Peruzzi) {2};
   \node [fill=col1] at (Bischeri) {1};
   \node [fill=col1] at (Ridolfi) {1};
   \node[v,label=20:\small $v$] at (Medici) {};
   \node[fill=col3] at (Salviati) {3};
   \node[fill=col3] at (Albizzi) {3};
   \node[fill=col3] at (Acciaiuoli) {3};
   \node[fill=col3] at (Tornabuoni) {3};
   \node[fill=col3] at (Barbadori) {3};
   \end{scope}
  \end{tikzpicture}
\quad
  \begin{tikzpicture}[scale=1.3,gray]
   \florentineee
    \begin{scope}[black,
   every node/.style={draw=black,circle, inner sep=1pt}]
   \draw [ultra thick](Guadagni)--(Lamberteschi);
   \node[fill=col1] at (Castellani) {1};
   \node[fill=col2] at (Peruzzi) {2};
   \node[fill=col1] at (Bischeri) {1};
   \node[fill=col1] at (Ridolfi) {1};
   \node [fill=col3] at (Salviati) {3};
   \node [fill=col3] at (Albizzi) {3};
   \node [fill=col3] at (Acciaiuoli) {3};
   \node [fill=col3] at (Tornabuoni) {3};
   \node [fill=col3] at (Barbadori) {3};
    \node [pattern=north east lines,pattern color=col3][preaction={fill=white}] at (Strozzi) {4};
    \node [pattern=north east lines,pattern color=col3][preaction={fill=white}] at (Medici) {4};
    \node [pattern=crosshatch dots,pattern color=col3][preaction={fill=white}] at (Guadagni) {5};
    \node [pattern=north east lines,pattern color=col3][preaction={fill=white}] at (Lamberteschi) {4};
    \node [pattern=north east lines,pattern color=col3][preaction={fill=white}] at (Ginori) {4};
    \node [pattern=north east lines,pattern color=col3][preaction={fill=white}] at (Pazzi) {4};
    \node [pattern=north east lines,pattern color=col3][preaction={fill=white}] at (Pucci) {4};
    \end{scope}
  \end{tikzpicture}
  \caption{\label{fig: Algorithm W}Execution of Algorithm W}
\end{figure}

Fig.~\ref{fig: Algorithm W} shows an execution of Algorithm W finding
a 5-colouring of the 16-vertex instance from Fig.~\ref{fig:
  cap}.

The running time is clearly bounded by $O(n+m)$.
  To analyse the number of colours, we first need to verify Step W2.
Since $G$ is 3-colourable, so is the subgraph induced by
$N(v)\cup\{v\}$.
Now, if $G[N(v)]$ requires 3 colours, then $G[N(v)\cup \{v\}]$
requires 4,
so $G[N(v)]$ is 2-colourable and therefore Step W2 is correct.
Note that Step~W2 can be run at most $O(\surd n)$ times, each using at most
two colours.
Step W3 expends another $\lceil\surd n\rceil$ colours according to
Algorithm G.

Algorithm W naturally extends to graphs with $\chi(G) >3$.
In this case, Step W2 calls Algorithm W recursively to colour
$(\chi(G)-1)$-colourable neighbourhoods.
The resulting algorithm uses $O(n^{1-1/(1-\chi(G))})$ colours.

\section{Recursion}

Recursion is a fundamental algorithmic design technique.
The idea is to reduce a problem to one or more simpler instances of
the same problem.

\subsection*{Contraction}

The oldest recursive construction for graph colouring expresses
the chromatic polynomial $P(G,q)$ and the chromatic number $\chi(G)$ in terms of edge-contractions:
For non-adjacent vertices $v$, $w$  and integer $q=0,1,\ldots,n$,
\begin{align*}
  P(G,q) & =  P(G+vw,q) + P (G/vw,q)\,,\\
  \chi(G)& = \min\{ \chi(G+vw) , \chi(G/vw)\} \,,
\end{align*}
see Chapter 3, Section 2.1.
These `addition--contraction' recurrences immediately imply a
recursive algorithm.
For instance,
\begin{align*}
P(\vcenter{\hbox{\begin{tikzpicture}[scale=.2]
  \node (0) [v] at (0,1)  {};
  \node (1) [v] at (0,-1) {};
  \node (2) [v] at (2,1)  {};
  \node (4) [v] at (2,-1)  {};
  \draw (0)--(1)--(2)--(0);
  \draw (1)--(4)--(0);
  \end{tikzpicture}}},q
)
&=
P(\vcenter{\hbox{\begin{tikzpicture}[scale=.2]
  \node (0) [v] at (0,1)  {};
  \node (1) [v] at (0,-1) {};
  \node (2) [v] at (2,1)  {};
  \node (4) [v] at (2,-1)  {};
  \draw (0)--(1)--(2)--(0);
  \draw (1)--(4)--(0);
  \draw (2)--(4);
 \end{tikzpicture}}},q
)
+
P(\vcenter{\hbox{\begin{tikzpicture}[scale=.25]
  \node (0) [v] at (0,1)  {};
  \node (1) [v] at (0,-1) {};
  \node (2) [v] at (1,0)  {};
   \draw (0)--(1)--(2)--(0);
   \end{tikzpicture}}},q
)\\
&= P(K_4,q)+P(K_3,q) = q(q-1)(q-2)\bigl((q-3)(q-4)+1\bigr) \,.
\end{align*}
Note that the graphs at the end of the recursion are complete.

For sparse graphs, it is more useful to express the same idea as a
`deletion--contraction' recurrence, which deletes and contracts edges until
the graph is empty:
\[ P(G,q) = P(G/e,q) - P (G-e,q)\qquad(e\in E)\,.\]

Many other graph
problems beside colouring can be expressed by a deletion--contraction
recurrence.
The most general graph invariant that can be defined in this fashion is the Tutte polynomial (see
\cite{BHKK-tutte} and \cite{HPR} for its algorithmic aspects).

The algorithm implied by these recursions is sometimes called \emph{Zykov's
algorithm} \cite{Z}.
Here is the deletion--contraction version.

\begin{algor}{C}{Contraction}{Given a graph $G$, this algorithm returns
    the sequence of coefficients
    $(a_0,a_1,\ldots,a_n)$ of the chromatic polynomial
    $P(G,q)=\sum_{i=0}^n a_iq^i$.}
  \item[C1] [Base] If $G$ has no edges then return the coefficients
    $(0,0,\ldots,0,1)$, corresponding to the polynomial $P(G,q)=
    q^n$.
  \item[C2] [Recursion] Pick an edge $e$ and construct the graphs $G'=G/e$ and
    $G''=G-e$.
    Call Algorithm C recursively to compute $P(G',q)$ and $P(G'',q)$ as
    sequences of coefficients $(a_0',a_1',\ldots, a_n')$ and
    $(a_0'',a_1'',\ldots, a_n'')$.
    Return $(a_0' - a_0'',a_1'-a_1'',\ldots, a_n' - a_n'')$, corresponding to the
    polynomial $P(G/e,q)-P(G-e,q)$. \qed
\end{algor}

To analyse the running time, let $T(r)$ be the number of
executions of Step~C1 for graphs with $n$ vertices and $m$ edges, where
$r=n+m$.
The two graphs constructed in Step~C2 have size $n-1+m-1=r-2$ and
$n+m-1=r-1$, respectively, so $T$ satisfies $T(r)=T(r-1)+T(r-2)$.
This is a well-known recurrence with solution $T(r) =O(\varphi^r)$, where
$\varphi=\frac{1}{2}(1+\surd 5)$ is the golden ratio.
Thus, Algorithm C requires $\varphi^{n+m}\poly(n)=O(1.619^{n+m})$ time.
A similar analysis for the algorithm implied by the deletion--addition
recursion gives $\varphi^{n+\overline{m}}\poly(n)$, where
$\overline{m}=\binom{n}{2}-m$ is the number of edges in the complement
of $G$.

These worst-case bounds are often very pessimistic.
They do not take into account that recurrences can be stopped as soon
as the graph is a tree (or some other easily recognized graph whose
chromatic polynomial is known as a closed formula), or that $P$
factorizes over connected components.
Moreover, we can use graph isomorphism heuristics and tabulation to
avoid some unnecessary recomputation of isomorphic subproblems (see \cite{HPR}).
Thus, Algorithm C is a more useful algorithm than its exponential
running time may indicate.

\subsection*{Vertex partitions and dynamic programming}

We turn to a different recurrence, which expresses
$\chi(G)$ in terms of induced subgraphs of $G$.
By taking $S$ to be a colour class of an optimal colouring of $G$, we
observe that every graph has an independent set of vertices $S$ for
which \( \chi(G) = 1+\chi(G-S)\).
Thus, we have
\begin{equation}\label{eq: dc}
  \chi(G) =  1 + \min \chi(G- S)\,,
\end{equation} where the minimum is taken over all non-empty
independent sets $S$ in $G$.

The recursive algorithm implied by \eqref{eq: dc} is too slow to be of
interest.
We expedite it using the fundamental algorithmic idea of \emph{dynamic
  programming}.
The central observation is that the subproblems $\chi(G-S)$ for
various vertex-subsets $S$ appearing in \eqref{eq: dc} are computed
over and over again.
It thus makes sense to store these $2^n$ values in a table when they
are first computed.
Subsequent evaluations can then be handled by consulting the table.

We express the resulting algorithm in a bottom-up fashion:

\begin{algor}{D}{Dynamic programming}{
  Given a graph $G$, this algorithm computes a table $T$ with  $T(W)=\chi(G[W])$, for each
  $W\subseteq V$.}
\item[D1] [Initialize] Construct a table with (initially undefined)
  entries $T(W)$ for each $W\subseteq V$.
  Set $T(\emptyset)=0$.
\item[D2] [Main loop]
  List all vertex-subsets $W_1,W_2,\ldots,W_{2^n}\subseteq V$ in
  non-decreasing order of their size.
  Do Step D3 for $W=W_2,W_3,\ldots, W_{2^n}$, then terminate.
\item[D3] [Determine $T(W)$]
  Set $T(W) = 1 + \min T(W\setminus S)$, where the minimum is taken
  over all non-empty independent sets $S$ in $G[W]$.
  \qed
\end{algor}

The ordering of subsets in the main loop D2 ensures that each set is
handled before any of its supersets.
In particular, all values $T(W\setminus S)$ needed in Step~D3 will have
been previously computed, so the algorithm is well defined.
The minimization in Step~D3 is implemented by iterating over all
$2^{|W|}$ subsets of $W$.
Thus, the total running time of Algorithm~D is within a polynomial
factor of
\begin{equation}
\label{eq: D time}
\sum_{W\subseteq V} 2^{|W|} = \sum_{k=0}^n \binom{n}{k} 2^k = 3^n\,.
\end{equation}

This rather straightforward application of dynamic programming already
provides the non-trivial insight that the chromatic number can be
computed in  time exponential in the number of vertices, rather than depending
exponentially on $m$, $\chi(G)$, or a superlinear function of $n$.

\subsection*{Maximal independent sets}
To pursue this idea a little further we notice that $S$ in \eqref{eq:
  dc} can be assumed to be a \emph{maximal} independent set -- that is,
not a proper subset of another independent set.
To see this, let $\col$ be an optimal colouring and consider the
colour class $S=\col^{-1}(1)$.
If $S$ is not maximal, then repeatedly pick a vertex $v$ that is not
adjacent to $S$, and set $\col(v)=1$.

By considering the disjoint union of $\frac{1}{3}k$ triangles, we see
that there  exist $k$-vertex graphs with $3^{k/3}$ maximal
independent sets.
It is known that this is also an upper bound, and that the maximal
independent sets can be enumerated within a polynomial factor of that
bound (see \cite{BronKerbosch}, \cite{MM} and \cite{TTT}).
We therefore have the following result:

\begin{thm}\label{thm: bk} The maximal independent sets of a graph on $k$ vertices
  can be listed in time $O(3^{k/3})$ and polynomial space.
\end{thm}

We can apply this idea to Algorithm~D.
The minimization in Step~D3 now takes the following form:
\begin{description}
\item[D3$\mathbf '$] [Determine $T(W)$]
  Set $T(W) = 1 + \min T(W\!\setminus\! S)$, where the minimum is taken
  over all maximal independent sets $S$ in $G[W]$.
\end{description}

Using Theorem~\ref{thm: bk} with $k=|W|$ for the minimization in Step~D3$'$,
the total running time of Algorithm~D comes within a polynomial factor
of
\[
\sum_{k=0}^n \binom{n}{k} 3^{k/3} = (1+3^{1/3})^n =O(2.443^n) \,.\]
For many years, this was the fastest known algorithm for the chromatic number.

\subsection*{3-colouring}
Of particular interest is the 3-colouring case.
Here, it makes more sense to let the outer loop iterate over all
maximal independent sets and check whether the complement is bipartite.

\begin{algor}{L}{Lawler's algorithm}{
  Given a graph $G$, this algorithm finds a $3$-colouring if one exists.}
  \item[L1] [Main loop] For each maximal independent set $S$ of $G$,
    do Step~L2.
  \item[L2]
    [Try $\col(S)=3$] Use Algorithm~B to find a colouring $\col\colon
    V\setminus S\rightarrow\{1,2\}$ of $G-S$ if one exists.
    In that case, extend $\col$ to all of $V$ by setting $\col(v)=3$
    for each $v\in S$, and terminate with $f$ as the result.
    \qed
\end{algor}

The running time of Algorithm L is dominated by the number of
executions of L2, which, according to Theorem~\ref{thm: bk}, is
$3^{n/3}$.
Thus, Algorithm L decides 3-colourability in time $3^{n/3}\poly(n)=O(1.442^n)$ and
polynomial space.

\bigskip

The use of maximal independent sets goes back to Christofides
\cite{Christofides}, while Algorithms D and L are due to Lawler
\cite{Law}.
A series of improvements to these ideas have further reduced these
running times.
At the time of writing, the best-known time bound for 3-colouring is
$O(1.329^n)$ by Beigel and Eppstein \cite{BE}.

\section{Subgraph expansion}

The \emph{Whitney expansion} \cite{Whitney} of the chromatic
polynomial is
\[ P(G,q) = \sum_{A\subseteq E} (-1)^{|A|} q^{k(A)}\,;\] see Chapter~3,
Section~2 for a proof.
It expresses the chromatic polynomial as an alternating sum of terms,
each of which depends on the number of connected components $k(A)$ of
the edge-subset $A\subseteq E$.
Determining $k(A)$ is a well-studied algorithmic graph problem, which
can be solved in time $O(n+m)$ (for example, by depth-first search).
Thus, the Whitney expansion can be evaluated in time $O(2^m(n+m))$.

\bigskip

A more recent expression (see \cite{BH08}) provides an expansion over
\emph{induced} subgraphs:

\begin{thm}
  For $W\subseteq V$, let $g(W)$ be the number of non-empty
  independent sets in $G[W]$.
  Then $G$ can be $q$-coloured if and only if
   \begin{equation}\label{eq: ie c}
 \sum_{W\subseteq V}
     (-1)^{|V\setminus W|} \bigl(g(W)\bigr)^q> 0\,.
   \end{equation}
\end{thm}

\begin{proof}
  For each $W\subseteq V$, the term $\bigl(g(W)\bigr)^q$ counts the
  number of ways of selecting $q$ non-empty independent sets $S_1,S_2,\ldots,
  S_q$, where $S_i\subseteq W$.
  For $U\subseteq V$, let $h(U)$ be the number of ways of selecting
  $q$ non-empty independent sets whose union is $U$.
  Then $(g(W))^q = \sum_{U\subseteq W} h(U)$, so
  \begin{align*} \sum_{W\subseteq V} (-1)^{|V\setminus W|} \bigl(g(W)\bigr)^q &=
  \sum_{W\subseteq V}  (-1)^{|V\setminus W|} \sum_{U\subseteq W} h(U)\\ &=
  \sum_{U\subseteq V} h(U) \sum_{W\supseteq U} (-1)^{|V\setminus
    W|} =  h(V)\,.
  \end{align*}
  For the last step, note that the inner sum (over $W$, with
  $U\subseteq W\subseteq V$) vanishes except when $U=V$, because there
  are as many odd-sized as even-sized sets sandwiched between
  different sets, by the principle of inclusion--exclusion.

  If $h(V)$ is non-zero, then there exist independent sets
  $S_1,S_2,\ldots,S_q$ whose union is $V$. These sets correspond to a
  colouring: associate a colour with the vertices in each set,
  breaking ties arbitrarily.
\end{proof}

For each $W\subseteq V$, we can compute the value $g(W)$ in time
$O(2^{|W|}m)$ by constructing each non-empty subset of $W$ and testing
it for independence.
Thus, the total running time for evaluating $\eqref{eq: ie c}$ is
within a polynomial factor of $3^n$, just as in the analysis
\eqref{eq: D time} for Algorithm D; however, the space requirement
here is only polynomial.
We can further reduce the running time to $O(2.247^n)$ by using
dedicated algorithms for evaluating $g(W)$ from the literature
(see \cite{BHK}).

If exponential space is available, we can do even better.
To that end, we first introduce a recurrence for $g$.
\begin{thm}\label{thm: g}
Let $W\subseteq V$. We have $g(\emptyset) = 0$, and, for every $v\in W$,
  \begin{equation}\label{eq: g}
    g(W)= g(W\setminus \{v\}) +
    g(W\setminus  N[v])+1\,.
  \end{equation}
\end{thm}

\begin{proof}
  Fix $v\in W$.
  The non-empty independent sets $S\subseteq W$ can be partitioned
  into two classes with $v\notin S$ and $v\in S$.
  In the first case, $S$ is a non-empty independent set with
  $S\subseteq W\setminus\{v\}$ and thus accounted for by the first
  term of \eqref{eq: g}.
  Consider the second case.
  Since $S$ contains $v$ and is independent, it contains no vertex
  from $N(v)$.
  Thus, $S$ is a non-empty independent set with $\{v\}\subseteq
  S\subseteq W\setminus N(v)$.
  The number of such sets is the same as the number of (not
  necessarily non-empty) independent sets $S'$ with $S'\subseteq
  W\setminus N[v]$, because of the bijective mapping $S\mapsto S'$
  where $S'=S\setminus\{v\}$.
  By induction, the number of such sets is $g(W\setminus N[v])+1$,
  where the `$+1$' term accounts for the empty set.
\end{proof}

This leads to the following algorithm, due to Björklund \emph{et al.} \cite{BHK}:

\begin{algor}{I}{Inclusion--exclusion}{
Given a graph $G$ and an integer $q\geq 1$, this algorithm determines whether $G$ can be
$q$-coloured.}
  \item[I1] [Tabulate $g$] Set $g(\emptyset) = 0$.
    For each non-empty subset $W\subseteq V$ in inclusion order, pick
    $v\in W$ and set
    \( g(W)= g(W\setminus \{v\}) +
    g(W\setminus N[v])+1\).
  \item[I2] [Evaluate \eqref{eq: ie c}]
    If $\sum_{W\subseteq V} (-1)^{|V\setminus W|} \bigl(g(W)\bigr)^q> 0$ output
    `yes', otherwise `no'.
    \qed
\end{algor}

\begin{figure}[ht]
\centerline{
  \raisebox{-1.3cm}{
  \begin{tikzpicture}
    \node (1) [v,label=right:$u$] at(1,0) {};
    \node (2) [v,label=left:$v$] at(0,0) {};
    \node (3) [v,label=below:$w$] at(.5,-.71) {};
    \node (4) [v,label=above:$x$] at(.5,.71) {};
    \draw (4)--(1)--(2)--(3)--(1);
  \end{tikzpicture}}\qquad
\small
  \begin{tabular}{ccrr}\toprule
    $W$& $g$ & $a_2$ & $a_3$\\ \midrule
    $\emptyset$ & 0 & 0 & 0\\
    $\{u\}$ & 1 & $-1$ & $-1$\\
    $\{v\}$ & 1 & $-1$ & $-1$\\
    $\{w\}$ & 1 & $-1$ & $-1$\\
    $\{x\}$ & 1 & $-1$ & $-1$\\
    $\{u,v\}$ & 2 & $4$ & $8$\\
    $\{u,w\}$ & 2 & $4$ & $8$\\
    $\{u,x\}$ & 2 & $4$ & $8$\\\bottomrule
  \end{tabular}
\quad
 \begin{tabular}{ccrr}\toprule
    $W$& $g$ & $a_2$ & $a_3$\\ \midrule
    $\{v,w\}$ & 2 & $4$ & $8$ \\
    $\{v,x\}$ & 3 & $9$ & $27$\\
    $\{w,x\}$ & 3 & $9$ & $27$ \\
    $\{u,v,w\}$ & 3 & $-9$ & $-27$\\
    $\{u,v,x\}$ & 4 & $-16$ & $-64$\\
    $\{u,w,x\}$ & 4 & $-16$ & $-64$\\
    $\{v,w,x\}$ & 5 & $-25$ & $-125$\\
    $V$ & 6 & $36$ & $216$\\\bottomrule
  \end{tabular}
}
\caption{\label{fig: Algorithm I}Execution of Algorithm I}
\end{figure}
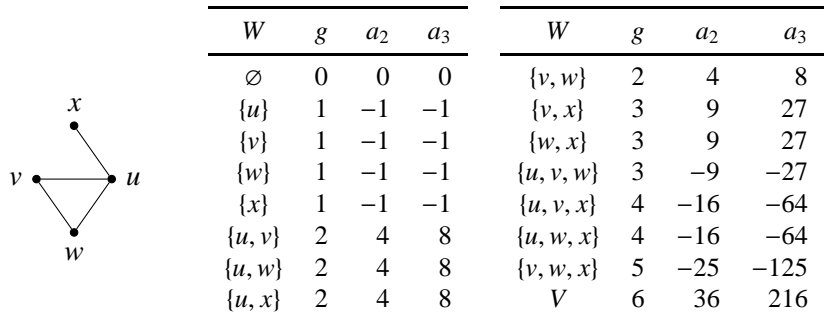

Both Steps I1 and I2 take time $2^n\poly(n)$, and the algorithm
requires a table with $2^n$ entries.
Fig.~\ref{fig: Algorithm I} shows the computations of Algorithm~I on a
small graph for $q=2$ and $q=3$, with $a_q(W)= (-1)^{|V\setminus W|} \bigl(g(W)\bigr)^q$.
The sum of the entries in column $a_2$ is $0$, so there is no 2-colouring.
The sum of the entries in column $a_3$ is $18$, so a 3-colouring exists.

\bigskip With slight modifications, Algorithm~I can be made to work
for other colouring problems such as the chromatic polynomial and
list-colouring, also in time and space $2^n\poly(n)$ (see \cite{BHK});
currently, this is the fastest known algorithm for these problems.
For the chromatic polynomial, the space requirement can be reduced to
$O(1.292^n)$, while maintaining the $2^n\poly(n)$ running time (see
\cite{BHKK-lin}).

\section{Local augmentation}

Sometimes, a non-optimal colouring can be improved by a local change
that recolours some vertices.
This general idea is the basis of many local search heuristics and
also several central theorems.

\subsection*{Kempe changes}

An important example, for edge-colouring, establishes Vizing's
theorem, $\Delta(G)\leq \chi'(G)\leq\Delta(G)+1$.
Chapter 5 gives a modern and more general presentation of the
underlying idea, and our focus in the present chapter is to make the
algorithm explicit.

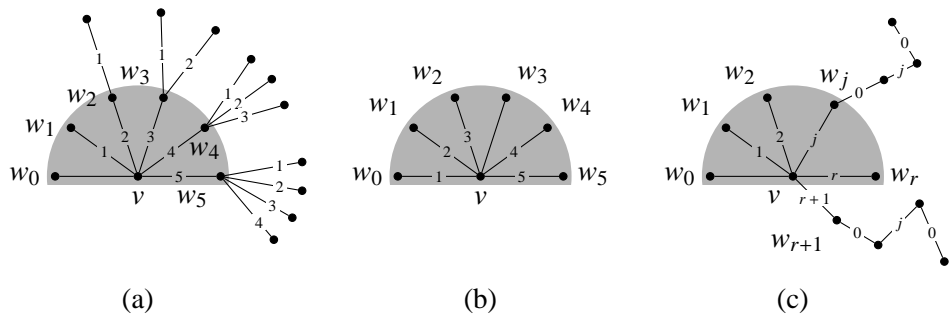
\begin{figure}[ht]
\tikzset{e/.style={fill=white,inner sep=1pt,font=\tiny}}
\begin{tikzpicture}[scale=1.1]
  \node at (0,-1.5) {(a)};
  \begin{pgfonlayer}{background}
  \clip (-1.1,-.1) rectangle (1.1,1.2);
  \fill [lightgray] (0,0) circle (1.1);
  \end{pgfonlayer}

  \node [v,label= below:$v$] (v) at (0,0) {};

  \node [v,label=left:$w_0$] (w0) at (180-0*36:1) {};

  \node [v,label=left:$w_1$] (w1) at (180-1*36:1) {};

  \node [v,label=left:$w_2$] (w2) at (180-2*36:1) {};
  \node [v] (w21) at (180-2*36:2) {};

  \node [v,label=120:$w_3$] (w3) at (180-3*36:1) {};
  \node [v] (w31) at (180-3*36+10:2) {};
  \node [v] (w32) at (180-3*36-10:2) {};

  \node [v,label=below:$w_4$] (w4) at (180-4*36:1) {};
  \node [v] (w41) at ($(w4)+(180-4*36+20:1)$) {};
  \node [v] (w42) at ($(w4)+(180-4*36:1)$) {};
  \node [v] (w43) at ($(w4)+(180-4*36-20:1)$) {};

  \node [v,label=235:$w_5$] (w5) at (180-5*36:1) {};
  \node [v] (w51) at ($(w5)+(180-5*36+10:1)$) {};
  \node [v] (w52) at ($(w5)+(180-5*36-10:1)$) {};
  \node [v] (w53) at ($(w5)+(180-5*36-30:1)$) {};
  \node [v] (w54) at ($(w5)+(180-5*36-50:1)$) {};

  \draw (v)--(w0);
  \draw (v) to node[e,fill=lightgray] {$1$} (w1);
  \draw (v) to node[e,fill=lightgray] {$2$} (w2);
  \draw (v) to node[e,fill=lightgray] {$3$} (w3);
  \draw (v) to node[e,fill=lightgray] {$4$} (w4);
  \draw (v) to node[e,fill=lightgray] {$5$} (w5);

  \draw  (w2) to node[e] {$1$} (w21);
  \draw  (w3) to node[e] {$1$} (w31);
  \draw  (w3) to node[e] {$2$} (w32);
  \draw  (w4) to node[e] {$1$} (w41);
  \draw  (w4) to node[e] {$2$} (w42);
  \draw  (w4) to node[e] {$3$} (w43);
  \draw  (w5) to node[e, near end] {$1$} (w51);
  \draw  (w5) to node[e,near end] {$2$} (w52);
  \draw  (w5) to node[e,near end] {$3$} (w53);
  \draw  (w5) to node[e,near end] {$4$} (w54);
\end{tikzpicture}
\quad
\begin{tikzpicture}[scale=1.1]
  \node at (0,-1.5) {(b)};
 \begin{pgfonlayer}{background}
  \clip (-1.1,-.1) rectangle (1.1,1.2);
  \fill [lightgray] (0,0) circle (1.1);
  \end{pgfonlayer}

  \node [v,label= below:$v$] (v) at (0,0) {};
  \node [v,label=180-0*36:$w_0$] (w0) at (180-0*36:1) {};
  \node [v,label=180-1*36:$w_1$] (w1) at (180-1*36:1) {};
  \node [v,label=180-2*36:$w_2$] (w2) at (180-2*36:1) {};
  \node [v,label=180-3*36:$w_3$] (w3) at (180-3*36:1) {};
  \node [v,label=180-4*36:$w_4$] (w4) at (180-4*36:1) {};
  \node [v,label=180-5*36:$w_5$] (w5) at (180-5*36:1) {};

  \draw (v)  to node[e,fill=lightgray] {$1$}(w0);
  \draw (v) to node[e,fill=lightgray] {$2$} (w1);
  \draw (v) to node[e,fill=lightgray] {$3$} (w2);
  \draw (v) to (w3);
  \draw (v) to node[e,fill=lightgray] {$4$} (w4);
  \draw (v) to node[e,fill=lightgray] {$5$} (w5);
\end{tikzpicture}
\quad
\begin{tikzpicture}[scale=1.1]
  \node at (0,-1.5) {(c)};
 \begin{pgfonlayer}{background}
  \clip (-1.1,-.1) rectangle (1.1,1.2);
  \fill [lightgray] (0,0) circle (1.1);
  \end{pgfonlayer}

  \node [v,label= 235:$v$] (v) at (0,0) {};
  \node [v,label=180-0*36:$w_0$] (w0) at (180-0*36:1) {};
  \node [v,label=180-1*36:$w_1$] (w1) at (180-1*36:1) {};
  \node [v,label=180-2*36:$w_2$] (w2) at (180-2*36:1) {};
  \node [v,label=above:$w_j$] (wj) at (60:1) {};
  \node [v] (P2) at ($(wj)+(.6,.3)$) {};
  \node [v] (P3) at ($(wj)+(1,.5)$) {};
  \node [v] (P4) at ($(wj)+(.7,1)$) {};
  \node [v,label=180-5*36:$w_r$] (wr) at (0:1) {};
  \node [v,label=235:$w_{r+1}$] (wr1) at (-45:.75) {};
  \node [v] (Q2) at ($(wr1)+(.5,-.3)$) {};
  \node [v] (Q3) at ($(wr1)+(1,.2)$) {};
  \node [v] (Q4) at ($(wr1)+(1.3,-.5)$) {};

  \draw (v)  to (w0);
  \draw (v) to node[e,fill=lightgray] {$1$} (w1);
  \draw (v) to node[e,fill=lightgray] {$2$} (w2);
  \draw (v) to node[e,fill=lightgray] {$j$} (wj);
  \draw (wj) to node[e] {$0$} (P2) to node[e] {$j$} (P3) to node[e]
  {$0$} (P4);
  \draw (wr1) to node[e] {$0$} (Q2) to node[e] {$j$} (Q3) to node[e]
  {$0$} (Q4);
  \draw (v) to node[e,fill=lightgray] {$r$} (wr);
  \draw  (v) to node[e] {$r+1$} (wr1);
\end{tikzpicture}
\caption{\label{fig: fan} (a) A fan\quad (b) Downshifting from $3$\quad (c) Step V7: colour $j$
  is free at $w_{r+1}$}
\end{figure}

A colour is \emph{free} at $v$ if it does not appear on an edge at $v$.
(We consider an edge-colouring with $\Delta(G)+1$ colours, so every
vertex has at least one free colour.)
A (Vizing) \emph{fan} around $v$ is a maximal set of edges $vw_0,vw_1,\ldots,vw_r$,
where $vw_0$ is not yet coloured and the other edges are coloured as
follows.
For $j=0,1,\ldots,r$, no colour is free at both $v$ and $w_j$.
For $j=1,2,\ldots,r$, the $j$th fan edge $vw_j$ has colour $j$ and the
colours appearing around $w_j$ include $1,2,\ldots,j$ but not $j+1$ (see Fig.~\ref{fig: fan}(a)).
Such a fan allows a recolouring by moving colours
as follows: remove the colour from $vw_j$ and set
  $\col(vw_0)=1,\allowbreak \col(vw_1)=2,\allowbreak\ldots,\allowbreak \col(vw_{j-1})=j$.
  This is called \emph{downshifting from $j$} (see Fig.~\ref{fig: fan}(b)).

\begin{algor}{V}{Vizing's algorithm}{Given a graph $G$, this algorithm finds an edge
    colouring with at most $\Delta(G)+1$ colours in time $O(nm)$.}
\item[V1]
  [Initialize]
  Order the edges arbitrarily $e_1,e_2,\ldots,e_m$.
  Let $i=0$.
\item[V2]
  [Extend colouring to next edge]
  Increment $i$. If $i=m+1$ then terminate. Otherwise, let $vw=e_i$.
\item[V3] [Easy case]
  If a colour $c$ is free at both $v$ and $w$, then set
  $\col(vw)=c$ and return to Step~V2.
\item[V4]
  [Find $w_0$ and $w_1$]
  Let $w_0=w$.
  Pick a free colour at $w_0$ and call it $1$.
  Let $vw_1$ be the edge incident with $v$ coloured $1$.
  (Such an edge exists because $1$ is not also free at $v$.)
\item[V5]
  [Find $w_2$]
  Pick a free colour at $w_1$ and call it $2$.
  If $2$ is also free at $v$, then set $\col(vw_0)=1$, $\col(vw_1)=2$, and
  return to Step~V2.
  Otherwise, let $vw_2$ be the edge incident with $v$ coloured $2$. Set $r=2$.
\item[V6]
  [Extend fan to $w_{r+1}$]
  Pick a free colour at $w_r$ and call it $r+1$.
  If $r+1$ is also free at $v$ then downshift from $r$, recolour
  $\col(vw_r)=c_{r+1}$ and return to Step~V2.
  Otherwise, let $vw_{r+1}$ be the edge incident with $v$ coloured $r+1$.
  If each colour $1,2,\ldots,r$ appears around $w_{r+1}$, then increment
  $r$ and repeat Step~V6.
\item[V7]
  [Build a $\{0,j\}$-path from $w_j$ or from $w_{r+1}$]
  Let $j\in\{1,2,\ldots, r\}$ be a free colour at $w_{r+1}$
  and let $0$ be a colour free at $v$ and different from $j$.
  Construct two maximal $\{0,j\}$-coloured paths $P_j$ and $P_{r+1}$
  from $w_j$ and $w_{r+1}$, respectively, by following edges of
  alternating colours $0,j,0,j,\ldots$ (see Fig.~\ref{fig: fan}(c)).
  (The paths cannot both end in $v$.)
  Let $k=j$ or $r+1$ so that $P_k$ does not end in $v$.
\item[V8] [Flip colours on  $P_k$] Recolour the edges on $P_k$ by exchanging $0$
  and $j$. Downshift from $k$, recolour $\col(vw_k)=0$, and return to Step~V2.\qed
\end{algor}

To see that this algorithm is correct, one needs to check that
the recolourings in Steps V6 and V8 are legal.
A careful analysis is given by Misra and Gries \cite{MiGr}.

For the running time, first note that Step V6 is repeated at most
$\deg v$ times, so the algorithm eventually has to leave that step.
The most time-consuming step is Step~V7; a $\{0,j\}$-path can be
constructed in time $O(n)$ if for each vertext we maintain a table of
incident edges indexed by colour.
Thus the total running time of Algorithm~V is $O(mn)$.

\medskip

Another example from this class of algorithms appears in the proof of
Brooks's theorem (see Chapter~2 and \cite{Brooks}), which relies on an algorithm that
follows Algorithm G but attempts to re-colour the vertices of
bichromatic components whenever a fresh colour is about to be
introduced.

\subsection*{Random changes}

There are many other graph colouring algorithms that fall under the
umbrella of local transformations.
Of particular interest are local search algorithms that recolour
individual vertices at random.
This idea defines a random process on the set of colourings called the
\emph{Glauber} or \emph{Metropolis} dynamics, or the natural Markov chain Monte
Carlo method.
The aim here is not merely to find a colouring (since
$q>4\Delta$, this would be easily done by Algorithm G), but to find a
colouring that is uniformly distributed among all $q$-colourings.

\begin{algor}{M}{Metropolis}{Given a graph $G$ with maximum degree
    $\Delta$ and a $q$-colouring $\col_0$ for $q> 4\Delta$, this algorithm finds a
    uniform random $q$-colouring $f_T$ in polynomial time.}
\item[M1]
  [Outer loop]
  Set $T=\lceil  qn\ln 2n/ (q-4\Delta)\rceil$. Do Step M2
  for $t=1,2,\ldots, T$, then terminate.
\item[M2]
  [Recolour a random vertex]
  Pick a vertex $v\in V$ and a colour $c\in \{1,2,\ldots,q\}$ uniformly at
  random.
  Set $\col_t=f_{t-1}$.
  If $c$ does not appear among $v$'s neighbours, then set $\col_t(v)=c$.
  \qed
\end{algor}

An initial colouring $f_0$ can be provided in polynomial time because $q>
\Delta+1$ -- for example, by Algorithm G.
To see that the choice of initial colouring $f_0$ has no influence on
the result $f_T$, we consider two different initial colourings $f_0$ and
$f_0'$ and execute Algorithm~M on both, using the same random choices
for $v$ and $c$ in each step.

Let $d_t=|\{\, v\colon f_t(v)\neq f_t'(v)\,\}|$ be the number
of \emph{disagreeing} vertices after $t$ executions of Step M2.
Each step can change only a single vertex, so
$|d_t-d_{t-1}|=1$, $0$, or $-1$.
We have $d_t= d_{t-1}+1$ only if $f_{t-1}(v)=f_{t-1}'(v)$ but
$f_t(v)\neq f_t'(v)$, so exactly one of the two processes rejects the
colour change.
In particular, $v$ must have a (disagreeing) neighbour $w$ with
$c=f_{t-1}(w)\neq f_{t-1}'(w)$ or $f_{t-1}(w)\neq f_{t-1}'(w)=c$.
There are $d_{t-1}$ choices for  $w$ and therefore $2\Delta
d_{t-1}$ choices for  $c$ and $v$.
Similarly, we have $d_t=d_{t-1} -1$ only if $f_{t-1}(v)\neq
f_{t-1}(v)$ and $c$ does not appear in $v$'s neighbourhood in either
$f_{t-1}$ or $f_{t-1}'$.
There are at least $(q-2\Delta)d_{t-1} $ such choices for $c$ and $v$.

Thus, the expected value of $d_t$ can be bounded as follows:
\begin{equation*}\mathbf E[d_t]\leq
  \mathbf E[d_{t-1}] + \frac{(q-2\Delta)\mathbf E[d_{t-1}]}{qn} -
\frac{2\Delta \mathbf E[d_{t-1}]}{qn} =
\mathbf E [d_{t-1}]\biggl(1-\frac{q-4\Delta}{qn}\biggr)\,.\end{equation*} Iterating this
argument and using  $d_0\leq n$, we have
\begin{equation*}
\mathbf E[d_T]\leq n\biggl(1-\frac{q-4\Delta}{qn}\biggr)^T \leq
  n\exp\biggl(-\frac{T(q-4\Delta)}{qn}\biggr) \leq n\exp(-\ln 2n)
  =\textstyle\frac{1}{2}\,.
\end{equation*} By Markov's inequality, and because $d_T$ is a
non-negative integer, we conclude that \[\Pr(f_T=f_T') = \Pr(d_T=0) \geq
1-\Pr(d_T\geq 1)\geq 1-\mathbf E[d_T] \geq \textstyle\frac{1}{2}\,.\]

\medskip
We content ourselves with this argument, which shows that the
process is `sufficiently random' in the sense of being memoryless.
Informally, we can convince ourselves that $f_T$ is uniformly
distributed because we can assume that $f_0'$ in the above argument
was sampled according to such a distribution.
This intuition can be formalized using standard coupling arguments for
Markov chains; our calculations above show that the `mixing time' of
Algorithm~M is $O(n\log n)$.

\medskip

Algorithm~M and its variants have been well studied, and the
analysis can be much improved (see the survey of Frieze and Vigoda \cite{FV}).
Randomized local search has wide appeal across disciplines, including
simulations in statistical physics and heuristic methods in
combinatorial optimization.

\section{Vector colouring}
\label{sec: SDP}

We now turn to a variant of vertex-colouring that is particularly
interesting from an algorithmic point of view.

\subsection*{Vector chromatic number}

Let $S^{d-1} = \{\,\mathbf x \in \mathbb R^d \colon \lVert\mathbf
x\rVert = 1\,\}$.
A \emph{vector $q$-colouring} in $d\leq n$ dimensions is a mapping
$x\colon V\rightarrow S^{d-1}$ from the vertex-set to the set of
$d$-dimensional unit vectors for which neighbouring vectors are `far
apart', in the sense that their scalar product satisfies
\[ \langle x(v), x(w)\rangle
\leq -\frac{1}{q-1}\,, \qquad \text{for $vw\in E$}.
\]
The smallest such number $q$ is called the \emph{vector chromatic number}
$\vec\chi(G)$, which need not be an integer.
For instance, the vertices of the 3-chromatic cycle graph $C_5$ can be laid out
on the unit circle
in the form of a pentagram
$\vcenter{\hbox{\begin{tikzpicture}
[every node/.style={fill,draw, circle, inner sep = .5pt},scale=.25]
\foreach \i in {0,1,2,3,4}  \node (\i)  at  (90+72*\i:1) {};
\draw (0)--(2)--(4)--(1)--(3)--(0);
\end{tikzpicture}}} $.
Then the angle between vectors corresponding to neighbouring vertices
is $\frac{4}{5}\pi$, corresponding to the scalar product $-1/(\surd
5-1)$, so $\vec\chi(C_5)\leq \surd 5<3$.

\begin{thm}\label{thm: sandwich} If $G$ has clique number $\omega(G)$,
  then $\omega(G)\leq \vec\chi(G)\leq \chi(G)$.
\end{thm}
\begin{proof}
  For the first inequality, let $W$ be a clique in $G$ of size
  $r=\omega(G)$ and consider a vector $q$-colouring $x$ of $G$.
  Let $\mathbf y= \sum_{v\in W} x(v)$.
  Then \[ 0 \leq \langle \mathbf y, \mathbf y \rangle \leq r\cdot
  1+r(r-1) \cdot \biggl(-\frac{1}{q-1}\biggr)\,,
\]
which implies that $r\leq q$.

For the second inequality,  place the vertices belonging to each
colour class at the corners of a $(q-1)$-dimensional simplex.
To be specific, let $f\colon V\rightarrow\{1,2,\ldots,q\}$ be an
optimal $q$-colouring and
define $x(v)=(x_1,x_2,\ldots, x_n)$ by
\begin{equation*}
x_i=
\begin{cases}
 \bigl((q-1)/q\bigr)^{1/2}\,, & \text{if $i=f(v)$}\,;\\
- \bigl(q(q-1)\bigr)^{-1/2}\,, & \text{if $i\neq f(v)$ and $i\leq q$}\,;\\
0\,, & \text{if $i> q$}\,.
\end{cases}
\end{equation*}
Then we have
\[\langle x(v), x(v)\rangle = \frac{q-1}{q} +
\frac{q-1}{q(q-1)} = 1\,,\]
and for $v$ and $w$ with $f(v)\neq f(w)$ we
have
\[\langle x(v),x(w)\rangle = 2 \biggl(\frac{q-1}{q}\biggr)^{1/2}
\biggl(- \biggl(\frac{q}{q-1}\biggr)^{1/2}\biggr) + \frac{q-2}{q(q-1)}
=-\frac{1}{q-1}\,.\]
Thus, $x$ is a vector $q$-colouring, so
$\vec\chi(G)$ is at most $q$.
\end{proof}

What makes vector colourings interesting from the algorithmic point of
view is that they can be found in polynomial time, at least
approximately, using algorithms based on semidefinite programming.
The details behind those constructions lie far outside the scope of
this chapter (see Gärtner and Matoušek \cite{GM}).

\begin{thm}\label{thm: sdp} Given a graph $G$ with
  $\vec\chi(G)=q$,  a vector $(q+\epsilon)$-colouring of $G$ can be
  found  in
  time polynomial in $n$ and $\log (1/\epsilon)$.
\end{thm}

For a graph with $\omega(G)=\chi(G)$, Theorem~\ref{thm: sandwich}
shows that the vector chromatic number equals the chromatic number.
In particular, it is an integer, and can be determined in polynomial
time using Theorem~\ref{thm: sdp} with $\epsilon <\frac{1}{2}$.
This shows that the chromatic numbers of perfect graphs can be
determined in polynomial time.
The theory behind this result counts as one of the highlights of
combinatorial optimization (see Grötschel, Lovász and
Schrijver \cite{GLS}).

How does the vector chromatic number behave for general graphs?
For $q=2$, the vectors have to point in exactly opposite directions.
In particular, there can be only two vectors for each connected
component, so vector 2-colouring is equivalent to 2-colouring.

\begin{figure}[ht]
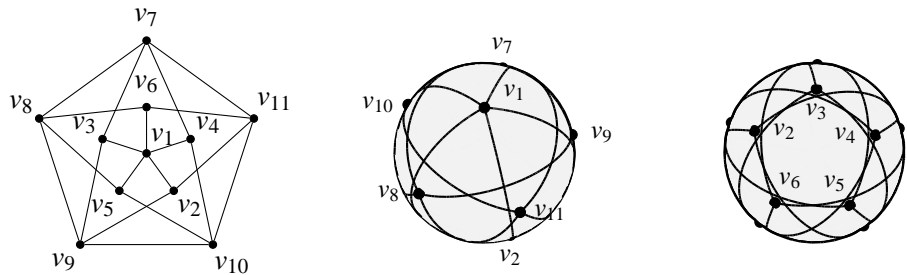

\centerline{
\begin{tikzpicture}[scale=1.5, every node/.style={v}]
\node   (1) at (0,0)         [label=45:$v_1$]         {};
\node   (7) at (90+72*0:1)   [label=90+72*0:$v_7$]    {};
\node   (8) at (90+72*1:1)   [label=170:$v_8$]    {};
\node   (9) at (90+72*2:1)   [label=90+72*2:$v_9$]    {};
\node  (10) at (90+72*3:1)   [label=90+72*3:$v_{10}$] {};
\node  (11) at (90+72*4:1)   [label=10:$v_{11}$] {};
\node   (2) at (90+72*0:.41) [label=90+72*0:$v_6$]  {};
\node   (3) at (90+72*1:.41) [label=170:$v_3$] {};
\node   (4) at (90+72*2:.41) [label=90+72*2:$v_5$] {};
\node   (5) at (90+72*3:.41) [label=90+72*3:$v_2$] {};
\node   (6) at (90+72*4:.41) [label=10:$v_4$] {};
\foreach \v in {2,3,4,5,6}  \draw (1)--(\v);
\draw (7)--(3)--(9)--(5)--(11);
\draw (2)--(8)--(4)--(10)--(6);
\draw (7)--(8)--(9)--(10)--(11);
\draw  (7) -- (11);
\draw (2) to  (11);
\draw (6) to  (7);
\end{tikzpicture}
\quad
\input{img/g-emb.pgf}\quad
\input{img/g-emb-S.pgf}
}
\caption{\label{fig: grotzsch clean} Left: the Grötzsch graph\qquad
 Middle and right: a vector 3-colouring}
\end{figure}

But already for $q=3$, the situation becomes more interesting, since
there exist vector 3-colourable graphs that are not 3-colourable.
For instance, the Grötzsch graph, the smallest triangle-free graph
with chromatic number 4, admits the vector 3-colouring shown in
Fig.~\ref{fig: grotzsch clean} as an embedding on the unit sphere.
More complicated constructions (that we cannot visualize) show that
there exist vector 3-colourable graphs with chromatic number at least
$n^{0.157}$ (see \cite{FLS} and \cite{KMS}).

\subsection*{Randomized rounding}

Even though the gap between $\vec\chi$ and $\chi$ can be large for
graphs in general, vector colouring turns out to be a useful starting
point for (standard) colouring.
The next algorithm, due to Karger, Motwani and Sudan \cite{KMS},
translates a vector colouring into a (standard) vertex-colouring using random
hyperplanes.

\begin{algor}{R}{Randomized rounding of vector colouring}{
  Given a $3$-chromatic graph $G$ with maximum degree $\Delta$,
  this algorithm finds a $q$-colouring in polynomial time, where the expected size of
  $q$ is $\mathbf
  E[q]=O(\Delta^{0.681} \log n)$.}
  \item[R1] [Vector colour] Set $\epsilon = 2\cdot 10^{-5}$ and
    compute a vector $(3+\epsilon)$-colouring $x$ of $G$ using
    semidefinite programming.
    Let $\alpha \geq \arccos(-1/(2+\epsilon))$ be the minimum
    angle in radians between neighbouring vertices.
  \item[R2] [Round] Set
 \[
    r=\lceil\log_{\pi/(\pi-\alpha)} (2\Delta)\rceil
    \]
    and construct $r$ random hyperplanes $H_1,H_2,\ldots, H_r$ in
    $\mathbb R^n$.
    For each vertex $v$, let $\col(v)$ be the binary number
    $b_rb_{r-1}\cdots b_1$, where $b_i=1$ if and only if $x(v)$ is on the positive
    side of the $i$th hyperplane $H_i$.
  \item[R3] [Handle monochromatic edges recursively] Iterate over all edges to
    find the set of monochromatic edges $M=\{\, vw\in E\colon
    \col(v)=\col(w)\,\}$.
    Recolour these vertices by running Algorithm R recursively on
    $G[M]$, with fresh colours. \qed
\end{algor}

\begin{figure}[ht]
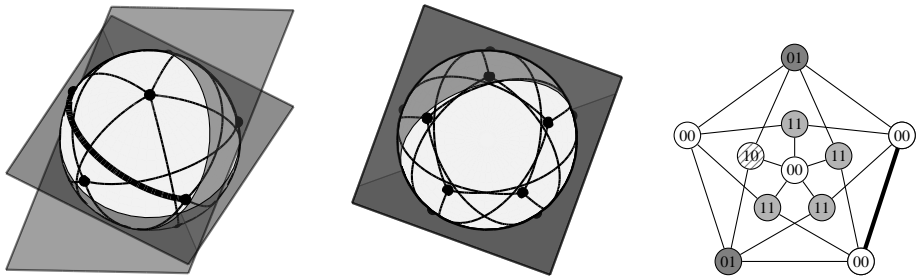

\centerline{
\input{img/g-emb-hyp-2.pgf}
\quad
\input{img/g-emb-hyp-1.pgf}
\quad
\begin{tikzpicture}[scale=1.5, every node/.style={draw, circle, inner sep=1pt}]
\node [fill=white]      (1) at  (0,0)        {\tiny 00};
\node [fill=gray]       (7) at  (90+72*0:1)  {\tiny 01};
\node [fill=white]      (8) at  (90+72*1:1)  {\tiny 00};
\node [fill=gray]       (9) at  (90+72*2:1)  {\tiny 01};
\node [fill=white]     (10) at  (90+72*3:1)  {\tiny 00};
\node [fill=white]     (11) at  (90+72*4:1)  {\tiny 00};
\node [fill=lightgray]  (2) at (90+72*0:.41) {\tiny 11};
\node [pattern=north east lines,pattern color=gray] (3) at (90+72*1:.41) {\tiny 10};
\node [fill=lightgray]  (4) at (90+72*2:.41) {\tiny 11};
\node [fill=lightgray]  (5) at (90+72*3:.41) {\tiny 11};
\node [fill=lightgray]  (6) at (90+72*4:.41) {\tiny 11};
\foreach \v in {2,3,4,5,6}  \draw (1)--(\v);
\draw (7)--(3)--(9)--(5)--(11);
\draw (2)--(8)--(4)--(10)--(6);
\draw (7)--(8)--(9)--(10)--(11)--(7);
\draw [line cap=rect,ultra thick] (10) -- (11);
\draw (2) to  (11);
\draw (6) to  (7);
\node [fill=white] (10)  at  (90+72*3:1) {\tiny 00};
\node [fill=white] (11) at  (90+72*4:1) {\tiny 00};
\end{tikzpicture}
}
\caption{\label{fig: grotzsch} Left and middle: two hyperplanes\qquad
 Right: the corresponding colouring}
\end{figure}

Figure~\ref{fig: grotzsch} illustrates the behaviour of Algorithm R on
the vector $3$-colouring of the Grötzsch graph from Fig.~\ref{fig:
  grotzsch clean}.
Two hyperplanes separate the vertices into four parts.
The resulting vertex-colouring with colours from $\{0,1\}^2$ is shown
to the right.
In this example, the set $M$ of monochromatic edges determined in Step
M3 contains only the single edge $v_{10}v_{11}$, drawn bold in the figure.

Algorithm R algorithm runs in polynomial time, because
Theorem~\ref{thm: sdp} ensures that Step~R1 can be performed in
polynomial time.

We proceed to analyze the size of the final colouring.
Step R2 uses the colours $\{0,1,\ldots, 2^{r-1}\}$, so the number of
colours used in each Step R2 is
\begin{equation}\label{eq: colsperround} 2^r \leq
 (2\Delta)^{-1/\log (\pi/(\pi-\alpha) )} < (2\Delta)^{0.631}\,,
\end{equation}
what is more difficult is to bound the total number of recursive
invocations.
To this end, we need to understand how fast the instance size,
determined by the size of $M$ in Step R3, shrinks.

Let $e$ be an edge whose endpoints received the vector colours $\mathbf
x$ and $\mathbf y$.
Elementary geometrical considerations establish the following result.

\begin{thm}\label{thm: angles}
  Let $\mathbf x , \mathbf y \in \mathbb R^d$ with angle $\varphi$ (in radians).
  A random hyperplane in $\mathbb R^d$ fails to separate $\mathbf x$ and
  $\mathbf y$ with probability  $1-\varphi/\pi$.
\end{thm}

The angle between the vectors $\mathbf x$ and $\mathbf y$ is at most $\alpha$.
(To gain some intuition of this, if we ignore the error term $\epsilon$,
Theorem~\ref{thm: angles} shows that $\mathbf x$ and $\mathbf y$ end up on the same
side of a random hyperplane with probability $1-\alpha/\pi \leq
1-\arccos(-\frac{1}{2})/\pi = 1-2\pi/3\pi = \frac{1}{3}.$)
The edge $e$ is monochromatic if all $r$ independent random hyperplanes
fail to separate $\mathbf x$ and $\mathbf y$ in Step~R2.
Thus,
\begin{equation*}
\Pr(e\in M) \leq (1-\alpha/\pi)^r \leq (\pi/(\pi-\alpha))^{-r} \leq 1/2\Delta\,.
\end{equation*} By
linearity of expectation, the expected size of $M$ is
\begin{equation*}
  \mathbf E [|M|]=\sum_{e\in E}\Pr(e\in M)
  \leq m/2\Delta\leq \textstyle\frac{1}{4}n\,.
\end{equation*}
Since each edge has two vertices, the expected number of vertices in
the recursive instance $G[M]$ is at most $\frac{1}{2}n$, and
in general, for $i>2$, the expected number of vertices $n_i$ in the
$i$th instance satisfies $n_i\leq \frac{1}{2}n_{i-1}$.
In particular, $n_t\leq 1$ after $t=O(\log n)$ rounds, at which point
the algorithm terminates.
With the bound \eqref{eq: colsperround} on the number of colours used
per round, we conclude that the total number of colours used is
$O(\Delta^{0.631} \log n)$ in expectation.

\bigskip In terms of $\Delta$, Algorithm R is much better than the
bound of $\Delta+1$ guaranteed by Algorithm G.
For an expression in terms of $n$, we are tempted to bound $\Delta$ by
$O(n)$, but that just shows that the number of colours is $O(n^{0.631}
\log n)$, which is worse than the $O(\surd n)$ colours from
Algorithm~W.

Instead, we employ a hybrid approach.
Run Steps W1 and W2 as long as the maximum degree of the graph $G$ is
larger than some threshold $d$, and
then colour the remaining graph using Algorithm R.
The number of colours used by the combined algorithm is of the order of
\( (2n/d) + (2d)^{0.631}\log n\), which is minimized around $d =
n^{1/1.631}$ with value $O(n^{0.387})$.

\bigskip Variants of Algorithm R for general $q$-colouring and with
intricate rounding schemes have been investigated further (see
Langberg's survey \cite{Langberg}).
The current best polynomial-time algorithm for colouring a
3-chromatic graph based on vector colouring uses $O(n^{0.208})$
colours, due to Chlamtac \cite{C}.

\section{Reductions}
\label{sec: reductions}

The algorithms in this chapter are summarized in Table~\ref{tab: Algorithms}.

\begin{table}[h!t]\small
\begin{tabular}{@{}l@{}llll@{}}\toprule
\multicolumn{2}{@{}l}{Algorithm} & Time & Problem\\ \midrule
{\bf B} & Bipartition &$O(n+m)$ & 2-colouring \\
{\bf C} & Contraction &  $O(1.619^{n+m})$ & $P(G,q)$ \\
{\bf D} & Dynamic programming & $3^n\poly(n)$ & $\chi(G)$ \\
{\bf G} & Greedy & $O(n+m)$ & $(\Delta(G)+1)$-colouring \\
{\bf I} & Inclusion--exclusion & $2^n\poly (n)$ & $\chi(G)$ \\
{\bf L} & Lawler's algorithm & $O(1.443^n)$ & 3-colouring \\
{\bf M\,\,} & Metropolis dynamics & $\poly(n)$ & random $q$-colouring ($q>4\Delta$) \\
{\bf P} & Palette restriction & $1.5^n\poly(n)$ & 3-colouring\\
{\bf R} & Rounded vector colouring & $\poly(n)$ & $O(\Delta^{0.681} \log
n)$-colouring for $\chi(G)=3$\\
{\bf V} & Vizing's algorithm & $O(mn)$ & edge  ($\Delta(G)+1$)-colouring \\
{\bf W} & Wigderson's algorithm & $O(n+m)$ & $O(\surd n)$-colouring for $\chi(G)=3$ \\
{\bf X} & Exhaustive search  & $q^n\poly(n)$ & $P(G,q)$ \\\bottomrule
\end{tabular}
\caption{\label{tab: Algorithms}
Algorithms discussed in this survey}
\end{table}

Not only do these algorithms achieve different running times and
quality guarantees, they also differ in which specific problem they
consider.
Let us now be more precise about the variants of the graph colouring
problem:

\begin{description}
\item[{\it Decision}] Given a graph $G$ and an integer $q$, decide whether $q$
  can be $q$-coloured.
\item[{\it Chromatic number}] Given a graph $G$, compute the chromatic
  number $\chi(G)$.
\item[{\it Construction}] Given a graph $G$ and an integer $q$, construct a
  $q$-colouring of $G$.
\item[{\it Counting}] Given a graph $G$ and an integer $q$, compute
  the number $P(G,q)$ of $q$-colourings of $G$.
\item[{\it Sampling}] Given a graph $G$ and an integer $q$, construct a random
  $q$-colouring of $G$.
\item[{\it Chromatic polynomial}] Given  a graph $G$, compute the chromatic
  polynomial -- that is, the coefficients of the integer polynomial $q\mapsto
  P(G,q)$.
\end{description}

Some of these problems are related by using fairly straightforward
reductions.
For example, the decision problem is easily solved using the
chromatic number by comparing $q$ with $\chi(G)$; conversely,
$\chi(G)$ can be determined by solving the decision problem for
$q=1,2,\ldots,n$.
It is also clear that if we can construct a $q$-colouring, then we can
decide that one exists.
What is perhaps less clear is the other direction.
This is seen by a self-reduction that follows the contraction algorithm,
Algorithm~C.

\medskip\noindent {\bf Reduction C}
(\emph{Constructing a colouring using a decision algorithm}). Suppose that we have an algorithm
that decides whether a given graph $G$ can be $q$-coloured.
If $G=K_n$ and $n\leq q$, give each vertex its own colour and
terminate.
Otherwise, select two non-adjacent vertices $v$ and $w$ in $G$.
If $G+vw$ cannot be $q$-coloured, then every $q$-colouring $\col$ of $G$
must have $\col(v)=\col(w)$.
Thus we can identify $v$ and $w$ and recursively find a $q$-colouring
for $G/vw$.
Otherwise, there exists a $q$-colouring of $G$ with $\col(v)\neq \col(w)$, so
we recursively find a colouring for $G+vw$.
\qed \medskip

Some of our algorithms work only for a specific fixed $q$, such as
Algorithm B for 2-colourability or Algorithm L for 3-colourability.
Clearly, they both reduce to the decision problem where $q$ is part
of the input.
But what about the other direction?
The answer turns out to depend strongly on $q$:
the decision problem reduces to 3-colorability, but not to
2-colorability.

\medskip
\noindent {\bf Reduction L}
(\emph{$q$-colouring using $3$-colouring}).
Given a graph $G= (V,E)$ and an integer $q$, this reduction constructs a graph $H$ that
is 3-colourable with colours $\{0,1,2\}$ if and only if $G$ is
$q$-colourable with colours $\{1,2,\ldots, q\}$.

First, to fix some colour names, the graph $H$ contains a triangle with
the vertices $0,1,2$.
We assume that vertex $i$ has colour $i$, for $i=0,1,2$.

For each vertex $v\in V$, the graph $H$ contains $2q$ vertices
$v_1,v_2,\ldots,v_q$ and $v_1',v_2',\ldots,v_q'$.
Our intuition is that the $v_i$s act as indicators for a colour in $G$
in the following sense: if $v_i$ has colour $1$ in $H$ then $v$
has colour $i$ in $G$.
The vertices are arranged as in Fig.~\ref{fig: reduction L}(a); the
right-most vertex is 1 or 2, depending on the parity of $q$.
\begin{figure}[ht]
\centerline{\begin{tikzpicture}[scale=.7]
\node (v1p)  [v,label=below:$v_1'$]  at (0,0) {};
\node (v1) [v,label=above:$v_{\vphantom{q}1}$] at (0,1) {};
\node (v2p)  [v,label=below:$v_2'$] at (1,0) {};
\node (v2) [v,label=above:$v_{\vphantom{q}2}$]  at (1,1) {};
\node (dots) at (2,0) {\small $\cdots$};
\node  at (2,1) {\small $\cdots$};
\node (vqp)  [v,label=below:$v_q'$] at (3,0) {};
\node (vq)  [v,label=above:$v_q$] at (3,1) {};
\node (2) [v,label=left:$2$] at (-1,0) {};
\node (0) [v,label=right:$2-q\pmod 2$] at (4,0) {};
\draw (v1p)--(v1);
\draw (v2p)--(v2);
\draw (vqp)--(vq);
\draw (v1p)--(v2p)--(dots)--(vqp);
\draw (2)--(v1); \draw (2)--(v2); \draw (2)--(v1p); \draw (2)--(vq);
\draw (vqp)--(0);
\node at (3,-2.5) {(a)};
\end{tikzpicture}
\qquad
\begin{tikzpicture}[scale=.7]
\node (w)  [v,label=-30:$w_i$]  at (-30:1cm) {};
\node (1) [v,label=above:$1$] at (90:1cm)  {};
\node (v)  [v,label=210:$v_i$] at (210:1cm) {};
\node (wp)  [v]  at (-30:.5cm) {};
\node (1p) [v] at (90:.5cm)  {};
\node (vp)  [v] at (210:.5cm) {};
\draw (w)--(wp)--(1p)--(vp)--(wp);
\draw (1p)--(1); \draw (vp)--(v);
\node at(0,-2.5) {(b)};
\end{tikzpicture}}
\vspace{-1em}
\caption{\label{fig: reduction L}\ }
\end{figure}
The vertices $v_1,v_2,\ldots,v_q$ are all adjacent to 2, and so must be
coloured 0 or 1.
Moreover, at least one of them must be coloured 1, since
otherwise, the colours for $v_1',v_2',\ldots,v_q'$ are forced to
alternate as $1,2,1,\ldots$, conflicting with the colour of the right-most
vertex.

Now consider an edge $vw$ in $G$.
Let $v_1,v_2,\ldots,v_q$ and $w_1,w_2,\ldots, w_q$ be the
corresponding `indicator' vertices in $H$.
For each colour $i=1,2,\ldots, q$, the vertices $v_i$ and $w_i$ are
connected by a `fresh' triangle as shown in Fig.~\ref{fig: reduction L}(b).
This ensures that $v_i$ and $w_i$ cannot both be 1.
In other words, $v$ and $w$ cannot have received the same colour.
\qed

\bigskip The above reduction, essentially due to Lovász \cite{Lov},
can easily be extended to a larger, fixed $q> 3$, because $G$ is
$q$-colourable if and only if $G$ with an added `apex' vertex
adjacent to all other vertices is $(q+1)$-colourable.
For instance, 4-colourability is not easier than 3-colourability for
general graphs.

Thus, all $q$-colouring problems for $q\geq 3$ are (in some sense)
equally difficult.
This is consistent with the fact that the case $q=2$ admits a very fast
algorithm (Algorithm~B), whereas none of the others does.

\bigskip

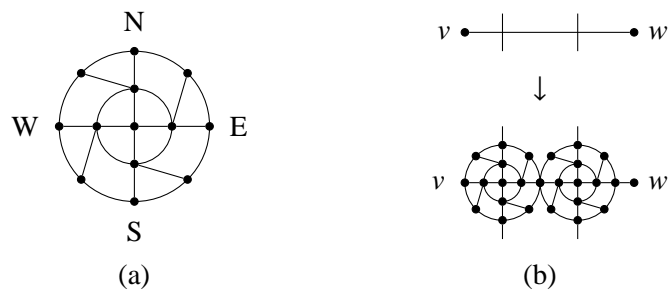
\begin{figure}[ht]
\centerline{
  \begin{tikzpicture}[scale=.5]
    \draw (0,0) circle (1);
    \draw (0,0) circle (2);
    \node (a) [v] at (0,0) {};
    \foreach \d in {0,90,180,270}
        \node (1\d) [v] at (\d:1) {};
    \foreach \d in {0,45, 90,...,315}
        \node (2\d) [v] at (\d:2) {};
     \node at (90:2) [label=north:N] {};
     \node at (180:2) [label=west:W] {};
     \node at (270:2) [label=south:S] {};
     \node at (0:2) [label=east:E] {};
    \draw (2180)--(1180)--(a)--(10)--(20);
    \draw (290)--(190)--(a)--(1270)--(2270);
    \draw  (190) to  (2135);
    \draw(1180) to (2225);
    \draw(1270) to (2315);
    \draw(10) to (245);
\node at (0,-4) {(a)};
  \end{tikzpicture}
\hspace{2cm}
  \begin{tikzpicture}[scale=.25]
    \node at (2,-5) {(b)};
    \node (vv) at (-2,8) [v,label=left:$v$] {};
    \node (ww) at (7,8) [v,label=right:$w$] {};
    \draw (vv)--(ww);
    \draw (0,7)--(0,9);
    \draw (4,7)--(4,9);
    \node at (2,5) {$\downarrow$};
    \node at (180:2) [label=left:$v$] {};
    \draw (0,0) circle (1);
    \draw (0,0) circle (2);
    \node (a) [v] at (0,0) {};
    \foreach \d in {0,90,180,270}
        \node (1\d) [v] at (\d:1) {};
    \foreach \d in {0,45, 90,...,315}
        \node (2\d) [v] at (\d:2) {};
    \draw (2180)--(1180)--(a)--(10)--(20);
    \draw (290)--(190)--(a)--(1270)--(2270);
    \draw  (190) to  (2135);
    \draw(1180) to (2225);
    \draw(1270) to (2315);
    \draw(10) to (245);
    \draw (4,0) circle (1);
    \draw (4,0) circle (2);
    \node (ta) [v] at (4,0) {};
    \foreach \d in {0,90,180,270}
        \node (t1\d) [v] at ([shift={(4,0)}]\d:1) {};
    \foreach \d in {0,45, 90,...,315}
        \node (t2\d) [v] at ([shift={(4,0)}] \d:2) {};
    \draw (t2180)--(t1180)--(ta)--(t10)--(t20);
    \draw (t290)--(t190)--(ta)--(t1270)--(t2270);
    \draw  (t190) to  (t2135);
    \draw(t1180) to (t2225);
    \draw(t1270) to (t2315);
    \draw(t10) to (t245);
    \node (w) at (7,0) [v,label=right:$w$] {};
    \draw(t20)--(w);
    \draw(290)-- +(0,1);
    \draw(2270)-- +(0,-1);
    \draw(t290)-- +(0,1);
    \draw(t2270)-- +(0,-1);
  \end{tikzpicture}
}
\caption{\label{fig: planar} A planarity gadget}
\end{figure}

Many constructions have been published that show the computational
difficulty of colouring for restricted classes of graphs.
We will sketch an interesting example due to Stockmeyer \cite{S}: the
restriction of the case $q=3$ to planar graphs.
Consider the subgraph in Fig.~\ref{fig: planar}(a), called a
\emph{planarity gadget}.
One can check that this subgraph has the property that every
3-colouring $\col$ satisfies $\col(\mathrm E)=\col(\mathrm W)$ and
$\col(\mathrm N)=\col(\mathrm S)$.
Moreover, every partial assignment $\col$ to $\{\mathrm N, \mathrm S,
\mathrm E, \mathrm W\}$ that satisfies $\col(\mathrm E)=\col(\mathrm
W)$ and $\col(\mathrm N)=\col(\mathrm S)$ can be extended to a
3-colouring of the entire subgraph.

The gadget is used to transform a given (non-planar) graph $G$ as
follows.
Draw $G$ in the plane and for each edge $vw$ replace each edge
intersection by the planarity gadget.
The outer vertices of neighbouring gadgets are identified, and $v$ is
identified with $\mathrm W$ in its neighbouring gadget (see
Fig.~\ref{fig: planar}(b)).
The resulting graph is planar, and it can be checked that it is
3-chromatic if and only if $G$ is 3-chromatic.
Thus, the restriction to planar instances does not make
3-colourability computationally easier.
Unlike the case for non-planar graphs, this construction cannot be generalized
to larger $q>3$, since the decision problem for planar graphs and
every $q\geq 4$ has answer `yes' because of the four-colour theorem.

\subsection*{Computational complexity}

The field of computational complexity relates algorithmic problems
from various domains to one another in order to establish a notion of
computational difficulty.
The chromatic number problem was one of the first to be analysed in
this fashion.
The following reduction, essentially from the seminal paper of Karp
\cite{Karp}, shows that computing the chromatic number is `hard for
the complexity class NP' by reducing from the NP-hard
satisfiability problem for Boolean formulas on conjunctive normal form (CNF).
This implies that all other problem in the class NP reduce to the
chromatic number.

The input to \emph{CNF-Satisfiability} is a Boolean formula consisting of $s$
clauses $C_1,C_2,\ldots,C_s$.
Each clause $C_j$ consists of a disjunction
$C_j=(l_{j1}\vee l_{j2}\vee \cdots\vee l_{jk})$ of literals.
Every literal is a variable $x_1,x_2,\ldots,x_r$ or its negation
$\overline{x}_1,\overline{x}_2,\ldots, \overline{x}_r$.
The problem is to find an assignment of  the variables  to `true' and
`false' that makes all clauses true.

\bigskip
\noindent {\bf Reduction K}
(\emph{Satisfiability using chromatic number}).
Given an instance $C_1,C_2,\allowbreak\ldots,\allowbreak C_s$ of CNF-Satisfiability over the
variables $x_1,x_2,\ldots, x_r$, this reduction constructs a graph $G$ on $3r+s+1$
vertices such that $G$ can be coloured with $r+1$ colours if and only
the instance is satisfiable.

\begin{figure}[ht]
\centerline{
\begin{tikzpicture}
\node (0) [circle,draw,inner sep=1pt,label=0] at (4.5,-1.25) {0};
\node (1) [circle,draw, fill=gray, inner sep=1pt,label=left:1]
          at (0,0) {1};
\node (2) [circle,draw, fill=lightgray,inner sep=1pt,label=left:2]
          at (0,-1.25) {2};
\node (3) [pattern=north east lines,pattern color=gray]
          [preaction={fill=white}][circle,draw,inner sep=1pt,label=left:3]
           at (0,-2.5) {3};
\node (v1)  at (2,.25)  [circle,draw,inner sep=1pt,fill=gray,label=right:$v_1$] {1};
\node (vn1) at (2,-.25) [circle,draw,inner sep=1pt,label=right:$\overline{v}_1$] {0};
\node (v2)  at (2,-1)   [circle,draw,inner sep=1pt,label=right:$v_2$] {0};
\node (vn2) at (2,-1.5) [circle,draw, fill=lightgray,inner sep=1pt,label=right:$\overline{v}_2$] {2};
\node (v3)  at (2,-2.25) [pattern=north east lines,pattern color=gray]
          [preaction={fill=white}]
[circle,draw,inner sep=1pt,label=right:$v_3$] {3};
\node (vn3) at (2,-2.75)[circle,draw,inner sep=1pt,label=right:$\overline{v}_3$] {0};
\draw (v1)--(vn1);
\draw (v1)--(2);\draw (v1)--(3);\draw (vn1)--(2);\draw (vn1)--(3);
\draw (v2)--(1);\draw (v2)--(3);\draw (vn2)--(1);\draw (vn2)--(3);
\draw (v2)--(vn2);
\draw (v3)--(1);\draw (v3)--(2);\draw (vn3)--(1);\draw (vn3)--(2);
\draw (v3)--(vn3);
\node (w1) at (3.25,-1.25) [fill=lightgray,circle,draw,inner sep=1pt,label=$w_1$] {2};
\foreach \v in {0,vn1,v2,v3} \draw (w1)--(\v);
\end{tikzpicture}
}
\caption{\label{fig: K} A 4-colouring instance corresponding to $C_1=(x_1\vee \overline{x}_2 \vee\overline{x}_3)$}
\end{figure}
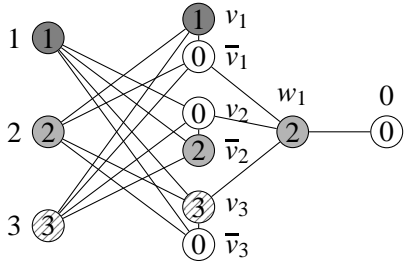

The graph $G$ contains a complete subgraph on $r+1$ vertices
$\{0,1,\ldots,r\}$.
In any colouring, these vertices receive different colours, say
$\col(i)=i$.
The intuition is that the colour $0$ represents `false', while the
other colours represent `true'.
For each variable $x_i$ $(1\leq i\leq r)$ the graph contains two
adjacent `literal' vertices $v_i$ and $\overline{v}_i$, both adjacent
to all `true colour' vertices $\{1,2,\ldots,r\}$ except $i$.
Thus, one of the two vertices $v_i,\overline{v}_i$ must be assigned
the `true' colour $i$, and the other must be coloured $0$.
The construction is completed with `clause' vertices $w_j$, one for
each clause $C_j$ $(1\leq j\leq s)$.
Let $x_{i_1}, x_{i_2}, \ldots, x_{i_k}$ be the variables appearing
(positively or negatively) in $C_j$.
Then $w_j$ is adjacent to $\{0,1,\ldots, r\}\setminus
\{i_1,i_2,\ldots, i_k\}$.
This ensures that only the `true' colours $\{i_1,i_2,\ldots, i_k\}$ are
available at $w_j$.
Furthermore, if $x_i$ appears positive in $C_j$, then $w_j$ is
adjacent to $\overline{v}_i$; if $x_i$ appears negated in $C_j$, then
$w_j$ is adjacent to $v_i$.
Figure~\ref{fig: K} shows the reduction for a small instance
consisting of just the clause $C_1=(x_1\vee \overline{x}_2
\vee\overline{x}_3)$ and a valid colouring corresponding to the
assignment $x_1=x_3=\text{true}$, $x_2=\text{false}$; the edges of the
clique on $\{0,1,2,3\}$ are not shown.
Thus, the only colours available to $w_j$ are those chosen by its
literals.
\qed

\subsection*{Edge-colouring}

A mapping $f\colon E\rightarrow\{1,2,\ldots, q\}$ is an edge-colouring
of $G$ if and only if it is a vertex-colouring of the line graph
$L(G)$ of $G$.
In particular, every vertex-colouring algorithm can be used as an edge-colouring algorithm by running it on $L(G)$.
For instance, Algorithm~I computes the chromatic index in time
$2^{m}\poly(n)$, which is the fastest currently known algorithm.
Similarly, Algorithm G finds an edge-colouring with $(2\Delta-1)$ colours, but this
is worse than Algorithm V.
In fact, since $\Delta\leq \chi'(G)\leq \Delta +1$, Algorithm V
determines the chromatic index within an additive error of 1.
However, deciding which of the two candidate values for $\chi'(G)$ is
correct is an NP-hard problem, as shown by Holyer \cite{Hol} for
$\chi'(G)=3$ and Leven and Galil \cite{LG} for $\chi'(G)>3$.

\subsection*{Approximating the chromatic number}

Algorithm V shows that the chromatic index can be very well
approximated.
In contrast, approximating the chromatic \emph{number} is much harder.
In particular, it is NP-hard to 4-colour a 3-chromatic graph
(see \cite{GuKh}).
This rules out an approximate vertex-colouring algorithm with a
performance guarantee as good as Algorithm V, but is far from explaining
why the considerable machinery behind, say, Algorithm R results only
in a colouring of size $n^c$ for 3-chromatic graphs.
The best currently known exponent is $c=0.204$ (see \cite{KaTh}).

For sufficiently large fixed $q$, it is NP-hard to find an
$\exp(\Omega(q^{1/3}))$-colouring for a $q$-colourable graph.
If $q$ is not fixed, even stronger hardness results are known.
We saw in Section~\ref{sec: SDP} that the polynomial-time computable
function $\vec\chi(G)$ is a lower bound on $\chi(G)$, even though the
gap can sometimes be large, say $\chi(G) \geq n^{0.157}\vec\chi(G)$
for some graphs.
Can we guarantee a corresponding upper bound for $\vec\chi$?
If not, maybe there is some other polynomial-time
computable function $g$ so that we can guarantee, for example,
$g(G)\leq \chi(G)\leq n^{0.999} g(G)$?
The answer turns out to be `no' under standard complexity-theoretic
assumptions: For every $\epsilon>0$, it is NP-hard to approximate
$\chi(G)$ within a factor $n^{1-\epsilon}$, as shown by Zuckerman \cite{Zuck}.

\subsection*{Counting}
The problem of counting the $q$-colourings is solved by evaluating
$P(G,q)$.
Conversely, because the chromatic polynomial has degree $n$, it can be
interpolated using Lagrangian interpolation from the values of the
counting problem at $q=0,1,\ldots, n$.
Moreover, note that $\chi(G)\geq q$ if and only if $P(G,q)>0$, so it is NP-hard
to count the number of $q$-colourings simply because the decision
problem is known to be hard.
In fact, the counting problem is hard for Valiant's counting class
$\#$P.

On the other hand, an important result in counting complexity
\cite{JVV} relates the estimation of the size of a finite set to the
problem of uniformly sampling from it.
In particular, a uniform sampler such as Algorithm~M serves as a
`fully polynomial randomized approximation scheme' (FPRAS) for the
number of colours.
Thus, provided that $q>4\Delta$, Algorithm~M can be used to compute a
value $g(G)$ for which $(1-\epsilon)g(G)\leq P(G,q) \leq (1+\epsilon)g(G)$
with high probability in time polynomial in $n$ and $1/\epsilon$ for
any $\epsilon>0$.
Much better bounds on $q$ are known (see the survey of Frieze and Vigoda \cite{FV}).
Without some bound on $q$, such an FPRAS is unlikely to exist because,
with $\epsilon = \frac{1}{2}$, it would constitute a randomized
algorithm for the decision problem and would therefore imply that all
of NP can be solved in randomized polynomial time.

\subsection*{Conclusion}

Together, the algorithms and reductions presented in this survey give
a picture of the computational aspects of graph colouring.
For instance, 2-colouring admits a polynomial time algorithm, while
3-colouring does not.
In the planar case, 4-colouring is trivial, but 3-colouring is not.
An almost optimal edge-colouring can be found in polynomial time, but
vertex-colouring is very difficult to approximate.
If $q$ is sufficiently large compared to $\Delta(G)$ then the set of
colourings can be sampled and approximately counted, but not counted
exactly.
Finally, even the computationally hard colouring problems admit
techniques that are much better than our initial Algorithm~X.

None of these insights is obvious from the definition of graph
colouring, so the algorithmic perspective on chromatic graph theory
has proved to be a fertile source of questions with interesting
answers.

\renewcommand{\bibname}{References}

\bigskip
\end{document}